\documentclass{aastex}
\usepackage{emulateapj5}
\usepackage{lscape}
\usepackage{onecolfloat}   

\newcounter{affil}
\newcommand{\hoggaffil}[2]{%
	\addtocounter{affil}{1}%
	\altaffiltext{\theaffil}{{#2}\label{#1}}}

\newcommand{\mpc}{\ensuremath{{\rm\,Mpc}}}

\newcommand{\hmpc}{\ensuremath{h^{-1}{\rm\,Mpc}}}
\newcommand{\ihmpc}{\ensuremath{h{\rm\,Mpc}^{-1}}}
\newcommand{\hmpcC}{\ensuremath{h^{-3}{\rm\,Mpc^3}}}
\newcommand{\hgpcC}{\ensuremath{h^{-3}{\rm\,Gpc^3}}}
\newcommand{\ihmpcC}{\ensuremath{h^3 {\rm\,Mpc}^{-3}}}
\newcommand{\kmsmpc}{\ensuremath{{\rm\ km\ s^{-1}\ Mpc^{-1}}}}

\newcommand{\Om}{\Omega_m}
\newcommand{\Omhh}{\Omega_mh^2}
\newcommand{\Obhh}{\Omega_bh^2}

\newcommand{\Veff}{V_{\rm eff}}

\newcommand{\bfr}{\ensuremath{\vec{r}}}

\newcommand{\beq}{\begin{equation}}
\newcommand{\eeq}{\end{equation}}
\newcommand{\beqa}{\begin{eqnarray}}
\newcommand{\eeqa}{\end{eqnarray}}

\newcommand{\tableskip}{\\[-6pt]}

\begin{document}
\twocolumn[

\submitted{Submitted to \textit{The Astrophysical Journal} 12/31/2004}
\lefthead{Baryon Acoustic Oscillations}

\title{Detection of the Baryon Acoustic Peak in the 
Large-Scale \\ Correlation Function of SDSS Luminous Red Galaxies}
\author{
Daniel J.\ Eisenstein\altaffilmark{\ref{Arizona},\ref{SF}},
Idit Zehavi\altaffilmark{\ref{Arizona}}, 
David W.\ Hogg\altaffilmark{\ref{NYU}}, 
Roman Scoccimarro\altaffilmark{\ref{NYU}}, 
Michael R.\ Blanton\altaffilmark{\ref{NYU}}, 
Robert C.\ Nichol\altaffilmark{\ref{Portsmouth}}, 
Ryan Scranton\altaffilmark{\ref{Pitt}},
Hee-Jong Seo\altaffilmark{\ref{Arizona}}, 
Max Tegmark\altaffilmark{\ref{Penn},\ref{MIT}}, 
Zheng Zheng\altaffilmark{\ref{IAS}},
Scott F.\ Anderson\altaffilmark{\ref{UW}},
Jim Annis\altaffilmark{\ref{FNAL}},
Neta Bahcall\altaffilmark{\ref{Princeton}},
Jon Brinkmann\altaffilmark{\ref{APO}},
Scott Burles\altaffilmark{\ref{MIT}},
Francisco J.\ Castander\altaffilmark{\ref{Barcelona}},
Andrew Connolly\altaffilmark{\ref{Pitt}},
Istvan Csabai\altaffilmark{\ref{Eotvos}},
Mamoru Doi\altaffilmark{\ref{TokyoDoi}},
Masataka Fukugita\altaffilmark{\ref{ICRR}},
Joshua A.\ Frieman\altaffilmark{\ref{FNAL},\ref{Chicago}},
Karl Glazebrook\altaffilmark{\ref{JHU}},
James E.\ Gunn\altaffilmark{\ref{Princeton}},
John S.\ Hendry\altaffilmark{\ref{FNAL}},
Gregory Hennessy\altaffilmark{\ref{Flagstaff}},
Zeljko Ivezi\'c\altaffilmark{\ref{UW}},
Stephen Kent\altaffilmark{\ref{FNAL}},
Gillian R.\ Knapp\altaffilmark{\ref{Princeton}},
Huan Lin\altaffilmark{\ref{FNAL}},
Yeong-Shang Loh\altaffilmark{\ref{Colorado}},
Robert H.\ Lupton\altaffilmark{\ref{Princeton}},
Bruce Margon\altaffilmark{\ref{STSCI}},
Timothy A.\ McKay\altaffilmark{\ref{Michigan}},
Avery Meiksin\altaffilmark{\ref{Edinburgh}},
Jeffery A.\ Munn\altaffilmark{\ref{Flagstaff}},
Adrian Pope\altaffilmark{\ref{JHU}},
Michael W.\ Richmond\altaffilmark{\ref{RIT}},
David Schlegel\altaffilmark{\ref{LBL}},
Donald P.\ Schneider\altaffilmark{\ref{PSU}},
Kazuhiro Shimasaku\altaffilmark{\ref{TokyoShim}},
Christopher Stoughton\altaffilmark{\ref{FNAL}},
Michael A.\ Strauss\altaffilmark{\ref{Princeton}},
Mark SubbaRao\altaffilmark{\ref{Chicago},\ref{Adler}},
Alexander S.\ Szalay\altaffilmark{\ref{JHU}},
Istv\'an Szapudi\altaffilmark{\ref{Hawaii}},
Douglas L.\ Tucker\altaffilmark{\ref{FNAL}},
Brian Yanny\altaffilmark{\ref{FNAL}},
\& Donald G.\ York\altaffilmark{\ref{Chicago}}
}

\begin{abstract}
We present the large-scale correlation function measured from a
spectroscopic sample of 46,748 luminous red galaxies from the Sloan
Digital Sky Survey.  The survey region covers $0.72\hgpcC$ over 3816
square degrees and $0.16<z<0.47$, making it the best sample yet for
the study of large-scale structure.  We find a well-detected peak in the correlation
function at $100\hmpc$ separation that is 
an excellent match to the predicted shape and location of the imprint
of the recombination-epoch acoustic oscillations on the low-redshift
clustering of matter.  This detection demonstrates the linear growth of
structure by gravitational instability between $z\approx 1000$ and the present
and confirms a firm prediction of the standard cosmological theory.
The acoustic peak provides a standard
ruler by which we can measure the ratio of the distances to $z=0.35$ and 
$z=1089$ to 4\% fractional accuracy and the absolute distance to $z=0.35$
to 5\% accuracy.  From the overall shape of the correlation function, we measure the matter density
$\Omega_mh^2$ to 8\% and find agreement with the value from 
cosmic microwave background (CMB) anisotropies.  
Independent of the constraints provided by the CMB acoustic
scale, we find $\Omega_m=0.273\pm0.025+0.123(1+w_0)+0.137\Omega_K$.
Including the CMB acoustic scale, we find that the spatial curvature is
$\Omega_K=-0.010\pm0.009$ if the dark energy is a cosmological constant.
More generally, our results provide a measurement of cosmological distance,
and hence an argument for dark energy,
based on a geometric method with the same simple physics as the microwave
background anisotropies.  
The standard cosmological model convincingly
passes these new and robust tests of its fundamental properties.
\end{abstract}

]

\hoggaffil{Arizona}{Steward Observatory, University of Arizona,
		933 N. Cherry Ave., Tucson, AZ 85121}
\hoggaffil{SF}{Alfred P.~Sloan Fellow}
\hoggaffil{NYU}{Center for Cosmology and Particle Physics, 
    Department of Physics, New York University,
    4 Washington Place, New York, NY 10003}
\hoggaffil{Portsmouth}{Institute of Cosmology and Gravitation, 
    Mercantile House, Hampshire Terrace, University of Portsmouth,
    Portsmouth, P01 2EG, UK}
\hoggaffil{Pitt}{Department of Physics and Astronomy, University of Pittsburgh,
    Pittsburgh, PA 15260}
\hoggaffil{Penn}{Department of Physics, University of Pennsylvania,
    Philadelphia, PA 19104}
\hoggaffil{MIT}{Department of Physics, Massachusetts Institute of Technology,
    Cambridge, MA 02139}
\hoggaffil{IAS}{School of Natural Sciences, Institute for Advanced Study, Princeton, NJ 08540}
\hoggaffil{UW}{Department of Astronomy, University of Washington, 
	    Box 351580, Seattle WA 98195-1580}
\hoggaffil{FNAL}{Fermilab National Accelerator Laboratory,
		P.O. Box 500, Batavia, IL 60510}
\hoggaffil{Princeton}{Princeton University Observatory, Peyton Hall,
		Princeton, NJ 08544}
\hoggaffil{APO}{Apache Point Observatory,
                P.O. Box 59, Sunspot, NM 88349}
\hoggaffil{Barcelona}{Institut d'Estudis Espacials de Catalunya/CSIC, Gran Capit\`a 2-4, E-08034 Barcelona, Spain}
\hoggaffil{Eotvos}{Department of Physics of Complex Systems, 
		E\"otv\"os University, P\'azm\'any P\'eter s\'et\'any 1, 
		H-1518 Budapest, Hungary}  
\hoggaffil{TokyoDoi}{Institute of Astronomy, School of Science, The University of Tokyo, 2-21-1 Osawa, Mitaka, Tokyo 181-0015, Japan}
\hoggaffil{ICRR}{Institute for Cosmic Ray Research, Univ.\ of Tokyo, Kashiwa 277-8582, Japan}
\hoggaffil{Chicago}{University of Chicago, Astronomy \& Astrophysics Center,
		5640 S. Ellis Ave., Chicago, IL 60637}
\hoggaffil{JHU}{Department of Physics and Astronomy,
                The Johns Hopkins University,
                3701 San Martin Drive, Baltimore, MD 21218}
\hoggaffil{Flagstaff}{United States Naval Observatory,
		 Flagstaff Station, P.O. Box 1149, Flagstaff, AZ  86002}
\hoggaffil{Colorado}{Center for Astrophysics and Space Astronomy,
Univ.\ of Colorado, Boulder, Colorado, 80803}
\hoggaffil{STSCI}{Space Telescope Science Institute, 3700 San Martin Drive, Baltimore, MD 21218}
\hoggaffil{Michigan}{Dept.\ of Physics, Univ.\ of Michigan, 
		Ann Arbor, MI 48109-1120} 
\hoggaffil{Edinburgh}{Institute for Astronomy,
		University of Edinburgh, Royal Observatory, 
		Blackford Hill, Edinburgh, EH9 3HJ, UK}
\hoggaffil{RIT}{Department of Physics, Rochester Institute of Technology, 
	85 Lomb Memorial Drive, Rochester, NY 14623-5603}
\hoggaffil{LBL}{Lawrence Berkeley National Lab, 1 Cyclotron Road, MS 50R-5032, Berkeley, CA 94720-8160}
\hoggaffil{PSU}{Department of Astronomy and Astrophysics,
                Pennsylvania State University, University Park, PA 16802}
\hoggaffil{TokyoShim}{Department of Astronomy, School of Science, The University of Tokyo, 7-3-1 Hongo, Bunkyo, Tokyo 113-0033, Japan}
\hoggaffil{Adler}{Adler Planetarium, 1300 S.\ Lake Shore Drive, Chicago, IL 60605}
\hoggaffil{Hawaii}{Institute for Astronomy, University of Hawaii, 2680 Woodlawn Drive, Honolulu, HI, 96822}

\section{Introduction}

In the last five years, the acoustic peaks in the cosmic microwave background (CMB)
anisotropy power spectrum have emerged as one of the strongest cosmological probes
\citep{Mil99,deB00,Han00,Hal01,Lee01,Net02,Benoit03,Pea03,Ben03}.
They measure the contents and curvature of the universe 
\citep{Jun96,Knox00,Lan01,Jaf01,Kno01,Efs02,Per02,Spe03,Teg03b}, 
demonstrate that the cosmic perturbations are generated early ($z\gg1000$) and are
dominantly adiabatic \citep{Hu96w,Hu96ww,Hu97,Pei03,Moo04},
and by their mere existence largely validate the
simple theory used to support their interpretation
\citep[for reviews, see][]{Hu97ss,Hu02}.

The acoustic peaks occur because the cosmological perturbations excite
sound waves in the relativistic plasma of the early universe
\citep{Pee70,Sun70,Bon84,Bon87,Hol89}.  The 
recombination to a neutral gas at redshift $z\approx1000$
abruptly decreases the sound speed
and effectively ends the wave
propagation.  In the time between the formation of the perturbations
and the epoch of recombination, modes of different wavelength can
complete different numbers of oscillation periods.  This translates
the characteristic time into a characteristic length scale and produces
a harmonic series of maxima and minima in the anisotropy power spectrum.

Because the universe has a significant fraction of baryons, cosmological
theory predicts that the acoustic oscillations in the plasma will also
be imprinted onto the late-time power spectrum of the non-relativistic
matter \citep{Pee70,Bon84,Hol89,Hu96,Eis98}. 
A simple way to understand this is to consider that from an initial point 
perturbation common to the dark matter and the baryons, the dark matter
perturbation grows in place while the baryonic perturbation is carried outward
in an expanding spherical wave \citep{Bas01,Bas02}.  
At recombination, this shell is roughly 150 Mpc in radius.  Afterwards,
the combined dark matter and baryon perturbation seeds the formation
of large-scale structure.  Because the central perturbation in the dark matter is
dominant compared to the baryonic shell, the acoustic feature is 
manifested as a small single spike in the correlation function at 150 Mpc separation.

The acoustic signatures in the large-scale clustering of galaxies 
yield three more opportunities to test the cosmological paradigm with the early-universe acoustic phenomenon: 
1) it would provide smoking-gun evidence for
our theory of gravitational clustering, notably the idea that large-scale
fluctuations grow by linear perturbation theory from $z\sim 1000$ to the present; 
2) it would give another confirmation of
the existence of dark matter at $z\sim1000$, since a fully baryonic
model produces an effect much larger than observed; 
and 3) it would provide
a characteristic and reasonably sharp length scale that can be measured
at a wide range of redshifts, thereby determining purely by geometry the
angular-diameter-distance-redshift relation and the evolution of the Hubble parameter
\citep{Eht98,Eis03}.
The last application can provide precise and robust constraints 
\citep{Bla03,Hu03,Lin03,Seo03,Ame04,Dol04,Mat04}  
on the acceleration of the expansion rate of the universe \citep{Rie98,Per99}.
The nature of the ``dark energy'' causing this acceleration is a complete
mystery at present \citep[for a review, see][]{Pad04}, 
but sorting between the various exotic explanations
will require superbly accurate data.
The acoustic peak method could provide a geometric complement
to the usual luminosity-distance methods such as those based on 
type Ia supernovae 
\citep[e.g.][]{Rie98,Per99,Kno03,Ton03,Rie04}.

Unfortunately, the acoustic features in the matter correlations are weak
(10\% contrast in the power spectrum) and on large scales.  This means
that one must survey very large volumes, of order 1$\hgpcC$, to detect
the signature
\citep{Teg97,Gol98,Eht98}.
Previous surveys 
have not found clean detections, due to sample size and (in some cases)
survey geometry.  
Early surveys \citep{Bro90,Lan96,Ein97} found anomalous peaks that, though
unlikely to be acoustic signatures \citep{Eis98c}, did
not reappear in larger surveys.
\citet{Per01} favored baryons at 2~$\sigma$ in a power spectrum analysis of data
from the 2dF Galaxy Redshift Survey \citep[hereafter 2dFGRS;][]{Col01}, but 
\citet{Teg02} did not recover the signal with 65\% of the same data.
\citet{Mil02} argued that due to smearing from the window function, the 2dFGRS result 
could only be due to the excess of power on large scales in baryonic models 
and not due to a detection of the oscillations themselves.
A spherical harmonic analysis of the full 2dFGRS \citep{Per04} 
did not find a significant baryon fraction.
\citet{Mil01} presented a ``possible detection'' (2.2~$\sigma$) from a combination
of three smaller surveys.  The analysis of the power spectrum of
the SDSS Main sample \citep{Teg03a,Teg03b} did not address the question 
explicitly but would not have been expected to detect the oscillations.
The analysis of the correlation function of the 2dFGRS \citep{Haw03}
and the SDSS Main sample \citep{Zeh04b} did not consider these scales.
Large quasar surveys \citep{Out03,Cro04a,Cro04b,Yah04} are too limited by shot noise to 
reach the required precision, although the high redshift does give leverage
on dark energy \citep{Out04} via the Alcock-Paczynski (1979) test.

In this paper, we present the large-scale correlation function of a
large spectroscopic sample of luminous, red galaxies (LRGs) from the
Sloan Digital Sky Survey \citep[SDSS;][]{Yor00}.  This sample covers 3816 square degrees
out to a redshift of $z=0.47$ with 46,748 galaxies.  While it contains
fewer galaxies than the 2dFGRS or the Main sample of the SDSS,
the LRG sample \citep{Eis01}
has been optimized for the study of structure on the largest scales
and as a result it is expected to significantly outperform those samples.
First results on intermediate and small-scale clustering of
the LRG sample were presented by \citet{Zeh04c} and \citet{Eis04}, and
the sample is also being used for the study of galaxy clusters and for
the evolution of massive galaxies \citep{Eis03b}.  Here we focus on
large scales and present the first clear detection of the acoustic peak
at late times.

The outline of the paper is as follows.  We introduce the SDSS and
the LRG sample in \S 2.   In \S 3, we present the 
LRG correlation function and its covariance matrix, along with a discussion
of tests we have performed.  In \S 4, we fit the correlation function to 
theoretical models and construct measurements on the cosmological distance
scale and cosmological parameters.  We conclude in \S 5 with a general
discussion of the results.  Readers wishing to focus on the results rather
than the details of the measurement may want to skip \S~3.2, \S~3.3, and \S~4.2. 

\section{The SDSS Luminous Red Galaxy Sample}

The SDSS \citep{Yor00,Sto02,Aba03,Aba04} is imaging $10^4$ square degrees of
high Galactic latitude sky in five passbands, $u$, $g$, $r$, $i$, and $z$
\citep{Fuk96,Gun98}.  Image processing \citep{Lup01,Sto02,Pie02,Ive04}
and calibration \citep{Hog01,Smi02} allow one to select galaxies,
quasars, and stars for follow-up spectroscopy with twin fiber-fed
double-spectrographs.  Targets are assigned to plug plates with a
tiling algorithm that ensures nearly complete samples \citep{Bla01t}; 
observing each plate generates 640 spectra covering 3800\AA\ to 9200\AA\ with
a resolution of 1800.  

We select galaxies for spectroscopy by two algorithms.
The primary sample \citep{Str02}, referred to here as the SDSS Main
sample, targets galaxies brighter than $r=17.77$.   The surface
density of such galaxies is about 90 per square degree,
and the median redshift is 0.10 with a tail out to $z\sim0.25$.  The LRG
algorithm \citep{Eis01} selects $\sim\!12$ additional galaxies
per square degree, using color-magnitude cuts in $g$, $r$, and $i$
to select galaxies to a Petrosian magnitude $r<19.5$ that are likely to be luminous
early-types at redshifts up to $\sim\!0.5$.  All fluxes are corrected
for extinction \citep{SFD} before use.  The selection is
extremely efficient, and the redshift success rate is very high.
A few additional galaxies (3 per square degree at $z>0.16$) matching the
rest-frame color and luminosity properties of the LRGs are extracted
from the SDSS Main sample; we refer to this combined set as the LRG
sample.  

For our clustering analysis, we use 46,748 luminous red galaxies over 
3816 square degrees and in the redshift range 0.16 to 0.47.  
The sky coverage of the sample is shown in \citet{Hog04} and is similar
to that of SDSS Data Release 3 \citep{Aba04}.
We require that the galaxies have rest-frame $g$-band absolute magnitudes
$-23.2<M_g<-21.2$ ($h=1$, $H_0=100h\kmsmpc$) where we have applied $k$ corrections and passively
evolved the galaxies to a fiducial redshift of $0.3$.  The resulting comoving
number density is close to constant out to $z=0.36$ (i.e., volume limited)
and drops thereafter; see Figure 1 of \citet{Zeh04c}.
The LRG sample is unusual as compared to flux-limited surveys because
it uses the same type of galaxy (luminous early-types) at all redshifts.

We model the radial and angular selection functions using the methods
in the appendix of \citet{Zeh04c}.  In brief, we build a model of the 
redshift distribution of the sample by integrating an empirical model of
the luminosity function and color distribution of the LRGs
with respect to the luminosity-color selection boundaries of the sample.
The model is smooth on small scales and includes the subtle interplay
of the color-redshift relation and the color selection boundaries.
To include slow evolutionary effects,
we force the model to the observed redshift histogram using
low-pass filtering. We will show in \S~\ref{sec:systematics} that
this filtering does not affect the correlation function on the
scales of interest.

The angular selection function is based on the spherical polygon
description of {\tt lss\_sample14} \citep{Bla04v}.  We model fiber
collisions and unobserved plates by the methods of \citet{Zeh04c}.
Regions with completeness below 60\% (due to unobserved plates) are
dropped entirely; the 3816 square degrees we use are highly complete.
The survey mask excludes regions around bright stars, but does not
otherwise model small-scale imperfections in the data.

The typical redshift of the sample is $z=0.35$.  The coordinate
distance to this redshift is $962\hmpc$ for $\Omega_m=0.3$, $\Omega_\Lambda=0.7$.
At this distance, $100\hmpc$ subtends $6.0^\circ$ on the sky and
corresponds to 0.04 in redshift.  
The LRG correlation length of $10\hmpc$ \citep{Zeh04c} subtends only $40'$ on the sky 
at $z=0.35$; viewed on this scale, the survey is far more bulk than boundary.

\begin{figure}[tb]
\plotone{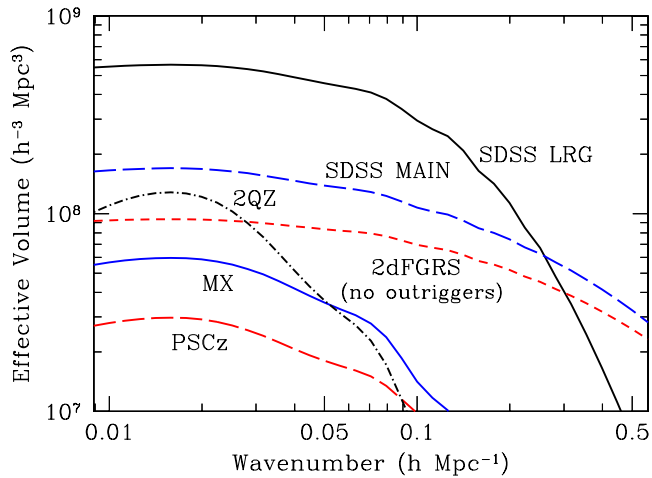}
\caption{\label{fig:Veff}
The effective volume (eq.~[\protect\ref{eq:Veff}]) as a function of wavenumber
for various large redshift surveys.  The effective volume is a rough guide to
the performance of a survey (errors scaling as $\Veff^{-1/2}$) but should
not be trusted to better than 30\%.
To facilitate comparison, we have assumed 3816 square degrees for the SDSS Main 
sample, the same area as the SDSS LRG sample presented in this paper
and similar to the area in Data Release 3.  This is about 50\% larger than 
the sample analyzed in \protect\citet{Teg03a}, which would be similar to the
curve for the full 2dF Galaxy Redshift Survey \protect\citep{Col03}.  
We have neglected the potential gains on very large scales from 
the 99 outrigger fields of the 2dFGRS.  
The other surveys are 
the MX survey of clusters \protect\citep{Mil01a}, 
the PSCz survey of galaxies \protect\citep{Sut99},  
and the 2QZ survey of quasars \protect\citep{Cro04a}. 
The SDSS DR3 quasar survey \protect\citep{Sch05} is similar in effective volume to the 2QZ.
For the amplitude of $P(k)$,
we have used $\sigma_8=1$ for 2QZ and PSCz and 3.6 for the MX survey.
We used $\sigma_8=1.8$ for SDSS LRG, SDSS Main, and the 2dFGRS; 
For the latter two, this value represents the amplitude of clustering of the
luminous galaxies at the surveys' edge; at lower redshift, the number density
is so high that the choice of $\sigma_8$ is irrelevant.
Reducing SDSS Main or 2dFGRS to $\sigma_8=1$, the value typical
of normal galaxies, decreases their $\Veff$ by 30\%.
}
\end{figure}

A common way to assess the statistical reach of a survey, including
the effects of shot noise, is by 
the effective volume \citep{Fel94,Teg97}
\beq\label{eq:Veff}
\Veff(k) = \int d^3r\;\left(n(\bfr)P(k)\over 1+n(\bfr)P(k)\right)^2
\eeq
where $n(\bfr)$ is the comoving number density of the sample at every location $\bfr$.
The effective volume is a function of the wavenumber $k$ via the power amplitude $P$.
For $P=10^4\hmpcC$ ($k\approx0.15\ihmpc$), we find $0.13\hgpcC$; 
for $P=4\times10^4\hmpcC$ ($k\approx0.05\ihmpc$), $0.38\hgpcC$; 
and for $P=10^5\hmpcC$ ($k\approx0.02\ihmpc$), $0.55\hgpcC$.  
The actual survey volume is $0.72\hgpcC$, so there are  
roughly 700 cubes of $100\hmpc$ size in the survey.
The relative sparseness of the LRG sample, $n\sim10^{-4}\ihmpcC$, 
is well suited to measuring power on large scales \citep{Kai86}.

In Figure \ref{fig:Veff}, we show a comparison of the effective volume of
the SDSS LRG survey to other published surveys.  While this calculation is
necessarily a rough ($\sim\!30\%$) predictor of statistical performance 
due to neglect of the exact survey boundary and
our detailed assumptions about the amplitude of the power spectrum and
the number density of objects for each survey, the SDSS LRG is clearly
the largest survey to date for studying the linear regime by a factor
of $\sim\!4$.  The LRG sample should therefore outperform these surveys
by a factor of 2 in fractional errors on large scales.  Note that
quasar surveys cover much more volume than even the LRG survey, but their
effective volumes are worse, even on large scales, due to shot noise.

\section{The redshift-space correlation function}

\subsection{Correlation function estimation}

In this paper, we analyze the large-scale clustering using the
two-point correlation function \citep{Pee80}.  
In recent years, the power spectrum has become
the common choice on large scales, as 
the power in different Fourier modes of the linear density field
is statistically independent in standard cosmology theories \citep{BBKS}.  
However, this advantage breaks down on small scales 
due to non-linear structure formation,
while on large scales, elaborate methods are required to recover the
statistical independence in the face of 
survey boundary effects \citep[for discussion, see][]{Teg98}.  
The power spectrum and correlation
function contain the same information in principle, as they are Fourier
transforms of one another.  
The property of the independence of different Fourier modes
is not lost in real space, but rather it is encoded into the off-diagonal
elements of the covariance matrix via a linear basis transformation.  
One must therefore accurately track the full
covariance matrix to use the correlation function properly, but this
is feasible.  An advantage of the correlation function is that, unlike
in the power spectrum, small-scale effects like shot noise and 
intra-halo astrophysics stay on small scales, well separated from the linear regime
fluctuations and acoustic effects.  

We compute the redshift-space correlation function using the Landy-Szalay
estimator \citep{Lan93}.  Random catalogs containing at least 16 times as 
many galaxies as the LRG sample were
constructed according to the radial and angular selection functions
described above.  We assume a flat cosmology with $\Omega_m=0.3$
and $\Omega_\Lambda=0.7$ when computing the correlation function.  We place
each data point in its comoving coordinate location based on its redshift
and compute the comoving
separation between two points using the vector difference.  We use
bins in separation of $4\hmpc$ from 10 to $30\hmpc$ and bins of $10\hmpc$ 
thereafter out to $180\hmpc$, for a total of 20 bins.  

We weight the sample using a scale-independent weighting that depends
on redshift.  When computing the correlation function, 
each galaxy and random point is weighted by $1/(1+n(z)P_{w})$ \citep{Fel94} 
where $n(z)$ is the comoving number density and $P_{w}=40,000\hmpcC$.  
We do not allow $P_{w}$ to change with scale so as to avoid scale-dependent
changes in the effective bias caused by differential changes in the sample redshift.  
Our choice of $P_w$ is close to optimal at $k\approx0.05\ihmpc$ and within
5\% of the optimal errors for all scales relevant to the acoustic oscillations 
($k\lesssim0.15\ihmpc$).
At $z<0.36$, $nP_{w}$ is about 4,
while $nP_{w}\approx1$ at $z=0.47$.  Our results do not qualitatively 
depend upon the value of $P_w$.

\begin{figure}[tb]
\plotone{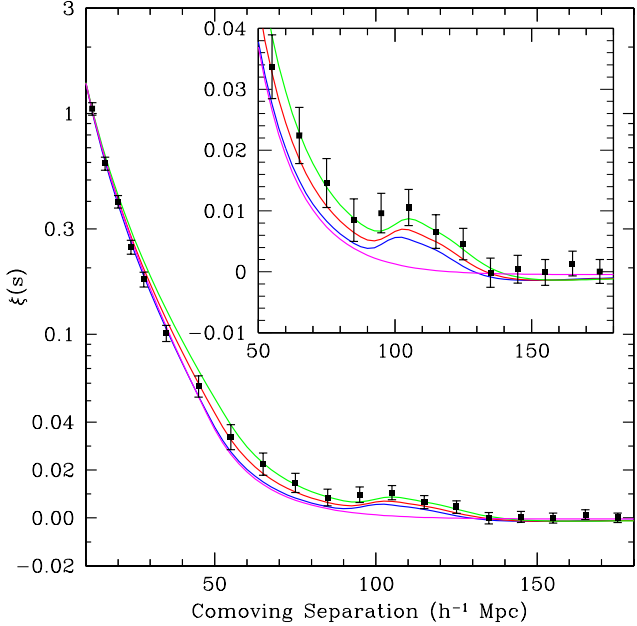}
\caption{\label{fig:xi}
The large-scale redshift-space correlation function of the SDSS LRG sample.
The error bars are from the diagonal elements of
the mock-catalog covariance matrix; however, the points are correlated.
Note that the vertical axis mixes logarithmic and linear scalings.
The inset shows an expanded view with a linear vertical axis.
The models are $\Omhh=0.12$ (top, green), 0.13 (red), and 0.14 (bottom with peak, blue), all
with $\Obhh=0.024$ and $n=0.98$ and with a mild non-linear prescription
folded in.  The magenta line shows a pure CDM model ($\Omhh=0.105$), 
which lacks the acoustic peak.
It is interesting to note that although the data appears higher than the
models, the covariance between the points is soft as regards overall shifts
in $\xi(s)$.  Subtracting 0.002 from $\xi(s)$ at all scales makes the plot look
cosmetically perfect, but changes the best-fit $\chi^2$ by only 1.3.
The bump at $100\hmpc$ scale, on the other hand, is statistically significant.
}
\end{figure}

Redshift distortions cause the redshift-space correlation function to vary
according to the angle between the separation vector and the line of sight.
To ease comparison to theory, we focus on the spherically averaged correlation
function.  Because of the boundary of the survey, the number of possible tangential 
separations is somewhat underrepresented compared to the number of possible line of sight
separations, particularly at very large scales.  To correct for this,
we compute the correlation functions in four angular bins.
The effects of redshift distortions
are obvious: large-separation correlations are smaller along the line of sight 
direction than along the tangential direction.  
We sum these four correlation functions in the proportions corresponding
to the fraction of the sphere included in the angular bin, thereby
recovering the spherically averaged redshift-space correlation function.
We have not yet explored the cosmological implications
of the anisotropy of the correlation function \citep{Mat03}.

\begin{figure}[tb]
\plotone{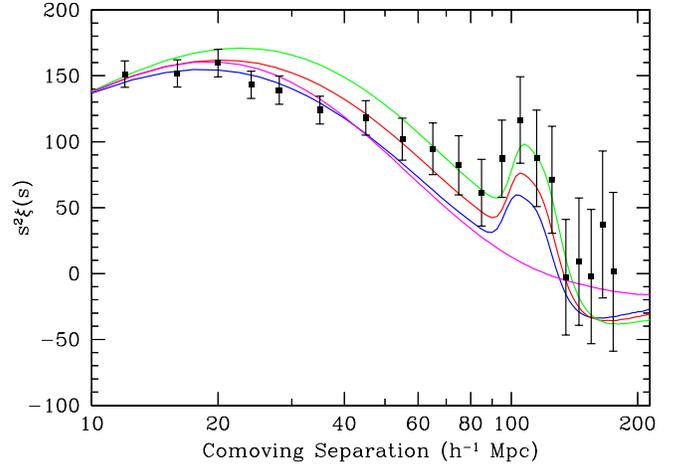}
\caption{\label{fig:xir2_jack}
As Figure \protect\ref{fig:xi}, but plotting the correlation function
times $s^2$.  This shows the variation of the peak at $20\hmpc$ scales
that is controlled by the redshift of equality (and hence by $\Omhh$).
Varying $\Omhh$ alters the amount of large-to-small scale correlation,
but boosting the large-scale correlations too much causes an inconsistency
at $30\hmpc$.  The pure CDM model (magenta) is actually close to the best-fit
due to the data points on intermediate scales.
}
\end{figure}

The resulting redshift-space correlation function is shown in 
Figure \ref{fig:xi}.  A more convenient view is in Figure \ref{fig:xir2_jack},
where we have multiplied by the square of the separation, so as
to flatten out the result.  The errors and overlaid models will be
discussed below.  The bump at $100\hmpc$ is the 
acoustic peak, to be described in \S \ref{sec:linear}.

The clustering bias of LRGs is known to be a strong function of luminosity 
\citep{Hog03,Eis04,Zeh04c}
and while the LRG sample is nearly volume-limited out to $z\sim0.36$,
the flux cut does produce a varying luminosity cut at higher redshifts.
If larger scale correlations were preferentially drawn from higher
redshift, we would have a differential bias \citep[see discussion in][]{Teg03a}.  
However, \citet{Zeh04c}
have studied the clustering amplitude in the two limiting cases,
namely the luminosity threshold at $z<0.36$ and that at $z=0.47$.
The differential bias between these two samples on large scales is 
modest, only 15\%.
We make a simple parameterization of the bias as a function of redshift
and then compute $b^2$ averaged as a function of scale over the pair 
counts in the random catalog.  The bias varies by less than 0.5\% as a function
of scale, and so we
conclude that there is no effect of a possible correlation of scale
with redshift.  This test also shows that the mean redshift as a function
of scale changes so little that variations in the clustering amplitude
at fixed luminosity as a function of redshift are negligible.

\subsection{Tests for systematic errors}\label{sec:systematics}

We have performed a number of tests searching for potential systematic errors
in our correlation function.
First, we have tested that the radial selection function is not introducing
features into the correlation function.  Our selection function involves
smoothing the observed histogram with a box-car smoothing of width
$\Delta z=0.07$.  This corresponds to reducing power in the purely
radial mode at $k=0.03\ihmpc$ by 50\%.  Purely radial power at $k=0.04$ $(0.02)\ihmpc$
is reduced by 13\% (86\%).  The effect of this suppression is negligible,
only $5\times10^{-4}$ ($10^{-4}$) on the correlation function at the $30$ (100) $\hmpc$ scale.
Simply put, purely radial modes are a small fraction of the total at these
wavelengths.
We find that an alternative radial selection function, in which
the redshifts of the random catalog are simply picked randomly
from the observed redshifts, produces a negligible change in the 
correlation function.  This of course corresponds to complete 
suppression of purely radial modes.

The selection of LRGs is highly sensitive to errors in the photometric
calibration of the $g$, $r$, and $i$ bands \citep{Eis01}.  We assess these by
making a detailed model of the distribution in color and luminosity
of the sample, including photometric errors, and then computing
the variation of the number of galaxies accepted at each redshift
with small variations in the LRG sample cuts.  A 1\% shift in the
$r-i$ color makes a 8-10\% change in number density; a 1\% shift in
the $g-r$ color makes a 5\% changes in number density out to $z=0.41$,
dropping thereafter; and a 1\% change in all magnitudes together
changes the number density by 2\% out to $z=0.36$, increasing to
3.6\% at $z=0.47$.  These variations are consistent with the 
changes in the observed redshift distribution when we 
move the selection boundaries to restrict the sample.
Such photometric calibration errors would cause anomalies in the correlation function 
as the square of the number density variations, 
as this noise source is uncorrelated with the 
true sky distribution of LRGs.

Assessments of calibration errors based on the color of the stellar locus
find only 1\% scatter in $g$, $r$, and $i$ \citep{Ive04}, which would
translate to about 0.02 in the correlation function.  However, the
situation is more favorable, because
the coherence scale of the calibration errors is limited
by the fact that the SDSS is calibrated in regions about $0.6^\circ$
wide and up to $15^\circ$ long.  This means that there are 20 independent
calibrations being applied to a given $6^\circ$ ($100\hmpc$) radius circular region.
Moreover, some of the calibration errors are even more localized, being
caused by small mischaracterizations of the point spread function and
errors in the flat field vectors early in the survey \citep{Sto02}.
Such errors will average down on larger scales even more quickly.

The photometric calibration of the SDSS has evolved slightly over time \citep{Aba03},
but of course the LRG selection was based on the calibrations at
the time of targeting.  We make our absolute magnitude cut on the 
latest (`uber') calibrations \citep{Bla04v,Fin04}, although this is only important at $z<0.36$
as the targeting flux cut limits the sample at higher redshift. 
We test whether changes in the calibrations might alter the
correlation function by creating a new random catalog, in which
the differences in each band between the target epoch photometry
and the `uber'-calibration values are mapped to angularly dependent 
redshift distributions using the number density derivatives presented above.
Using this random catalog makes negligible differences in the 
correlation function on all scales.  This is not surprising: while an rms
error of 2\% in $r-i$ would give rise to a 0.04 excess in the correlation
function, scales large enough that this would matter have many independent
calibrations contributing to them.
On the other hand, the `uber'-calibration of the survey does not necessarily
fix all calibration problems, particularly variations within a single
survey run, and so this test cannot rule out an arbitrary calibration
problem.  

We continue our search for calibration errors by breaking the survey
into 10 radial pieces and measuring the cross-correlations between the
non-adjacent slabs.  Calibration errors would produce 
significant correlations on large angular scales.  Some cross-correlations 
have amplitudes of 2-3\%, but many others don't, suggesting that
this is simply noise.  We also take the full matrix
of cross-correlations at a given separation and
attempt to model it 
(minus the diagonal and first off-diagonal elements) 
as an outer product of vector with itself, as
would be appropriate if it were dominated by a single type of 
radial perturbation, but we do not find plausible or stable vectors,
again indicative of noise.  Hence, we conclude that systematic errors in $\xi(r)$
due to calibration must be below 0.01.

It is important to note that calibration errors in the SDSS produce
large-angle correlations only along the scan direction.  Even if errors
were noticeably large, they would not produce narrow features such as
that seen at the $100 h^{-1}$ Mpc scale for two reasons.  First, the projections from 
the three-dimensional sphere to one-dimension strips on the sky necessarily
means that a given angular scale maps to a wide range of three-dimensional separations.
Second, the comoving angular diameter distance used to translate angles into 
transverse separations varies by a factor of three from $z=0.16$ to 0.47,
so that a preferred angle would not map to a narrow range of physical scales.
We therefore expect that calibration errors would appear as a smooth anomalous correlation,
rolling off towards large scales.

\begin{figure}[tb]
\plotone{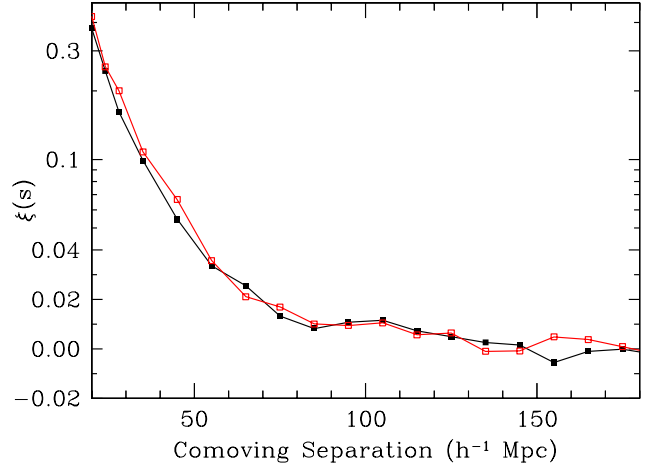}
\caption{\label{fig:xir2_redshift}
The correlation function for two different redshift slices,
$0.16<z<0.36$ (filled squares, black) and $0.36<z<0.47$ (open squares, red).  
The latter is somewhat noisier, but the two are quite similar and both
show evidence for the acoustic peak.  
Note that the vertical axis mixes logarithmic and linear scalings.
}
\end{figure}

Breaking the sample into two redshift slices above and below $z=0.36$
yields similar correlation functions (Fig.~\ref{fig:xir2_redshift}).
Errors in calibration or in the radial selection function would likely
enter the two redshift slices in different manners, but we see no sign
of this.  In particular, the bump in the correlation function appears in
both slices.  While this could in principle give additional leverage on
the cosmological distance scale, we have not pursued this in this paper.

\pagebreak
\subsection{Covariance Matrix}\label{sec:covar}

Because of the large number of separation bins and the large scales being
studied, it is infeasible to use jackknife sampling to construct a 
covariance matrix for $\xi(r)$.  Instead, we use a large set of mock catalogs to 
construct a covariance matrix and then test that matrix with a smaller
number of jackknife samples.

Our mock catalogs are constructed using PTHalos \citep{Sco02} with a halo
occupation model \citep{Ma00,Sel00,Pea00,Sco01,Coo02,Ber03}
that matches the observed amplitude of clustering of LRGs \citep{Zeh04c}.
PTHalos distributes dark matter halos according to extended Press-Schechter
theory conditioned by the large-scale density field from
second-order Lagrangian perturbation theory.  This generates
density distributions that fully include Gaussian linear theory,
but also include second-order clustering, redshift distortions, and
the small-scale halo structure that dominates the non-Gaussian signal.
We use a cosmology of $\Omega_m=0.3$, $\Omega_\Lambda=0.7$, $h=0.7$ for
the mock catalogs.
We subsample the catalogs so as to match the comoving density of LRGs as a function of redshift exactly,
but do not attempt to include the small redshift dependence in the
amplitude of the bias.  Our catalogs match the angular geometry of the
survey except in some fine details involving less than 1\% of the area.
Two simulation boxes, each $(1250\hmpc)^2\times 2500\hmpc$, one for the 
North Galactic Cap and the other for the South, 
are pasted together in each catalog.  These two regions
of the survey are very well separated in space, so this division is harmless.
We generate 1278 mock catalogs with independent initial conditions, compute the correlation function in
each, and compute the covariance matrix from the variations between them.

The resulting matrix shows considerable correlations between neighboring
bins, but with an enhanced diagonal due to shot noise.  The
power on scales below $10\hmpc$ creates significant correlation between
neighboring scales.  A curious aspect is that on large scales, where
our bins are $10\hmpc$ wide, the reduced inverse covariance matrix
is quite close to tri-diagonal; the first off-diagonal is typically
around 0.4, but the second and subsequent are typically a few percent.
Such matrices correspond to exponential decays of the off-diagonal
correlations \citep{Ryb94}.

We test the covariance matrix in several ways.  
First, as described in \S \ref{sec:basic}, the best-fit cosmological model has $\chi^2 = 16.1$ on 17 degrees
of freedom ($p=0.52$), indicating that the covariances are of the correct scale.
Next, we subdivide the survey into ten large, compact subregions and look
at the variations between jack-knifed samples (i.e., computing $\xi$ while 
excluding one region at a time).  
The variations of $\xi(r)$ between the ten jackknife samples matches
the diagonal elements of the covariance matrix to within 10\%.  We then
test the off-diagonal terms by asking whether the jack-knife residuals
vary appropriately relative to the covariance matrix.
If one defines the residual $\Delta\xi_j$ between the $j$th jackknife sample
and the mean, then the construction $\sum_{ab}\Delta\xi_j(r_a)C^{-1}_{ab}
\Delta\xi_j(r_b)$, where $C$ is the covariance matrix and the sums are
over the 20 radial bins, should have a mean (averaged over the ten jackknife sample)
of about $20/(10-1)\approx2.2$.  The mean value is $2.7\pm0.5$, in reasonable agreement.

We get similar results for cosmological parameters when using a
covariance matrix that is based on constructing the Gaussian approximation
\citep{Fel94,Teg97} of independent modes in Fourier space.  We use the effective 
volume at each wavenumber, plus an extra shot noise term to represent non-Gaussian
halos, and then rotate this matrix from the diagonal Fourier basis into
the real-space basis.  This matrix gives reasonable $\chi^2$ values, satisfies
the jackknife tests, and gives similar values and errors on cosmological parameters.

Finally, we test our results for cosmological parameters by using the following
hybrid scheme.  We use the mock-catalog covariance matrix to find the
best-fit cosmological model for each of the ten jackknife samples, and then
use the rms of the ten best-fit parameter sets to determine the errors.
This means that we are using the mock catalogs to weight the $\xi(r)$ measurements, but
relying on the jackknifed samples to actually determine the variance on 
the cosmological parameters.
We will quote the numerical results in \S~\ref{sec:basic}, but here we note that
the resulting constraints on the acoustic scale and matter density match
the errors inferred from the fitting with the mock-catalog covariance
matrix.  Hence, we conclude that our covariance matrix is generating correct
results for our model fitting and cosmological parameter estimates.

\section{Constraints on Cosmological Models}

\subsection{Linear theory}
\label{sec:linear}

Given the value of the matter density $\Omhh$, the baryon density $\Obhh$,
the spectral tilt $n$, and a possible neutrino mass, adiabatic cold dark
matter (CDM) models predict the linear matter power spectrum (and correlation function)
up to an amplitude factor.  There are two primary physical scales at work
\citep[for discussions, see][]{Hu97,Eis98,Hu02}.  
First,
the clustering of CDM is suppressed on scales small enough 
to have been traversed by the neutrinos and/or photons during the
radiation-dominated period of the universe.
This introduces the characteristic turnover in the CDM power spectrum
at the scale of the horizon at matter-radiation equality.  This length
scales as $(\Omhh)^{-1}$ (assuming the standard cosmic neutrino background).
Second, the acoustic oscillations have a characteristic scale known as the 
sound horizon, which is the comoving distance that a sound wave can travel
between the big bang and recombination.  This depends both on the expansion
history of the early universe and on the sound speed in the plasma; for models
close to the concordance cosmology, this length scales as 
$(\Omhh)^{-0.25}(\Obhh)^{-0.08}$ \citep{Hu04}.  The spectral tilt and massive neutrinos
tilt the low-redshift power spectrum \citep{Bon83}, but don't otherwise change these 
two scales.  Importantly, given $\Omhh$ and $\Obhh$, the two physical scales
are known in absolute units, with no factors of $H_0^{-1}$.

We use CMBfast \citep{Sel96z,Zal98,Zal00} to compute the linear power spectra, which we
convert to correlation functions with a Fourier transform.  It is important
to note that the harmonic series of acoustic peaks found in the power 
spectrum transform to a single peak in the correlation function \citep{Mat04}.  
The decreasing envelope of the higher harmonics, due to Silk damping \citep{Sil68}
or non-linear 
gravity \citep{Mei99}, corresponds to a broadening of the single peak.  
This broadening decreases the accuracy with which one can measure the centroid of the peak,
equivalent to the degradation of acoustic scale measurements caused by the disappearance of 
higher harmonics in the power spectrum \citep{Seo03}.

Examples of the model correlation functions are shown in Figures
\ref{fig:xi} and \ref{fig:xir2_jack}, each with $\Obhh=0.024$ and $h=0.7$
but with three values of $\Omhh$.  Higher values of $\Omhh$ correspond
to earlier epochs of matter-radiation equality, which increase the 
amount of small-scale power compared to large, which in turn decreases
the correlations on large scales when holding the small-scale amplitude fixed.
The acoustic scale, on the other hand, depends only weakly on $\Omhh$.

\subsection{Non-linear corrections}
\label{sec:nonlin}

The precision of the LRG correlations is such that we cannot rely
entirely on linear theory even at $r>10\hmpc$ (wavenumbers $k<0.1\ihmpc$).  
Non-linear gravity \citep{Mei99}, 
redshift distortions \citep{Kai87,Ham98,Sco04},
and scale-dependent bias all enter at a subtle level.  We address
these by applying
corrections derived from N-body simulations and the \citet{Smi03} {\tt halofit} method.

\citet{Seo04} present a set of 51 N-body simulations of the WMAP
best-fit cosmology \citep{Spe03}. 
Each simulation has $256^3$ particles and is $512\hmpc$ in size.  The outputs
at $z=0.3$ were analyzed to find their power spectra and correlation
functions, in both real and redshift-space, including simple halo bias.
This provides an accurate description of the non-linear gravitational
and redshift distortion effects on large scales.

\begin{figure}[tb]
\plotone{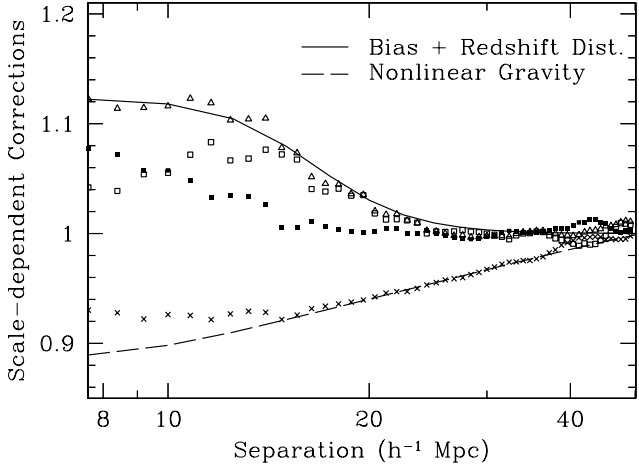}
\caption{\label{fig:corrections}
Scale-dependent corrections derived from 51 N-body simulations, each $512\hmpc$ comoving with $256^3$
particles \protect\citep{Seo04}.  The crosses show the
ratio between the non-linear matter correlation function and the linear
correlation function; the dashed line is the model we use from \protect\citet{Smi03}.
The solid points are the ratio between the 
biased correlation function (using a simple halo mass cut) to the 
non-linear matter correlation function.  The open squares are the
ratio of the biased redshift-space correlation function to the biased
real-space correlation function, after removing the large-scale
asymptotic value \citep{Kai87}, which we simply fold into 
the correlation amplitude parameter.  The open triangles show the product
of these two effects, and the solid line is our fit to this product.
These corrections are of order 10\% 
at $10\hmpc$ separations and decrease quickly on larger scales.
In addition to these corrections, we mimic the erasure of the 
small-scale acoustic oscillations
in the power spectrum by using a smoothed cross-over at $k=0.14\ihmpc$
between the CMBfast linear power spectrum and the no-wiggle form from Eisenstein
\& Hu (1998).
}
\end{figure}

For a general cosmology, we begin from the CMBfast linear power spectrum.
We next correct for the erasure of the higher acoustic peak that 
occurs due to mode-coupling \citep{Mei99}.
We do this by generating the ``no-wiggle'' approximation from \citet{Eis98}, 
which matches the overall shape of the linear power spectrum but with the 
acoustic oscillations edited out, and then smoothly interpolating between
the linear spectrum and the approximate one.  We use a Gaussian $\exp[-(ka)^2]$,
with a scale $a=7\hmpc$ chosen to approximately match the suppression
of the oscillations seen in the power spectrum of the N-body simulations.  

We next use the \citet{Smi03} package to compute the alterations in power from
non-linear gravitational collapse.  We use a model with $\Gamma=0.162$
and $\sigma_8=0.85$.  This model has the same shape for the power
spectrum on small scales as the WMAP best-fit cosmology and is therefore
a reasonable zero-baryon model with which to compute.  We find the
quotient of the non-linear and linear power spectra for this model
and multiply that ratio onto our baryonic power spectrum.  We then
Fourier transform the power spectrum to generate the real-space 
correlation function.  The accuracy of this correction is shown in 
the bottom of Figure \ref{fig:corrections}: the crosses are the 
ratio in the N-body simulations between the $z=0.3$ and $z=49$ 
correlation functions, while the dashed line is the \citet{Smi03}
derived correction.

Next we correct for redshift distortions and halo bias.  We use
the N-body simulations to find
the ratio of the redshift-space biased correlation function to 
the real-space matter correlation function for a halo mass threshold
that approximately matches the observed LRG clustering amplitude.
This ratio approaches an asymptotic value on large scales that
we simply include into the large-scale clustering bias.
After removing this asymptotic value, the
remaining piece of the ratio is only 10\% at $r=10\hmpc$.  We fit
this to a simple smooth function and then multiply the model
correlation functions by this fit.  Figure \ref{fig:corrections}
shows the N-body results for the ratio of redshift-space $\xi$ to 
real-space $\xi$ for the halos and the ratio of the real-space
halo correlation function to that of the matter.  The solid line
shows our fit to the product.  Tripling the mass threshold of
the bias model increases the correction by only 30\% (i.e., 1.12 
to 1.16 on small scales).

All of the corrections in Figure \ref{fig:corrections} are clearly
small, only 10\% at $r>10\hmpc$.  
This is because
$10\hmpc$ separations are larger than any virialized halo
and are only affected by the extremes of the finger of God redshift distortions. 
While our methods are not perfect, they are
plausibly matched to the allowed cosmology and to the bias of the 
LRGs.  As such, we believe that the corrections should be accurate to a few
percent, which is sufficient for our purposes.  Similarly, while
we have derived the corrections for a single cosmological model, 
the data constrain the allowed cosmologies enough that variations
in the corrections will be smaller than our tolerances.
For example, increasing $\sigma_8$ to 1.0 for the halofit calculation
changes the corrections at $r>10\hmpc$ by less than 2\%.

We stress that while galaxy clustering bias does routinely
affect large-scale clustering (obviously so in the LRG sample, with bias
$b\approx2$), it is very implausible that it would mimic the acoustic
signature, as this would require galaxy formation physics to have a strong
preferred scale at $100h^{-1}$ Mpc.  Galaxy formation prescriptions that
involve only small-scale physics, such as that involving dark matter halos
or even moderate scale radiation transport, necessarily produce smooth
effects on large scales \citep{Col93,Fry93,Sch98}.  Even long-range effects
that might be invoked would need to affect $100\hmpc$ scales differently from
80 or 130.  Our detection of the acoustic peak cannot reasonably
be explained as an illusion of galaxy formation physics.

\subsection{Measurements of the acoustic and equality scales}
\label{sec:basic}

The observed LRG correlation function could differ from that of the correct
cosmological model in amplitude, because of clustering bias and uncertain
growth functions, and in scale, because
we may have used an incorrect cosmology in converting from redshift
into distance.  Our goal is to use the comparison between observations
and theory to infer the correct distance scale.

Note that in principle a change in the cosmological model 
would change the distances differently
for different redshifts, requiring us to recompute the correlation
function for each model choice.  In practice, the changes are small enough
and the redshifts close enough that we treat the variation as a single dilation
in scale \citep[similar to][]{Bla03}.
This would be a superb approximation at low redshift, where all
distances behave inversely with the Hubble constant.  By $z=0.35$, the effects
of cosmological acceleration are beginning to enter.  However, we have checked explicitly
that our single-scale approximation is good enough for $\Omega_m$ between 0.2 and 0.4.  
Relative to our fiducial scale at $z=0.35$,
the change in distance across the redshift range $0.16<z<0.47$ is only 3\% peak to
peak for $\Omega_m=0.2$ compared to 0.3, and even these variations largely
cancel around the $z=0.35$ midpoint where we will quote our cosmological constraints.

The other error in our one-scale-parameter approximation is to treat the 
line-of-sight dilation 
equivalently to the transverse dilation.  In truth, the Hubble parameter
changes differently from the angular diameter distance (the Alcock-Paczynski (1979)
effect).  For small deviations from $\Omega_m=0.3$, $\Omega_\Lambda=0.7$, the 
change in the Hubble parameter at $z=0.35$ is about half of that of
the angular diameter distance.  We model this by treating the dilation scale
as the cube root of the product of the radial dilation times the square
of the transverse dilation.  In other words, we define
\begin{equation}\label{eq:D}
D_V(z) = \left[ D_M(z)^2 {cz\over H(z)}\right]^{1/3}
\end{equation}
where $H(z)$ is the Hubble parameter and $D_M(z)$ is the comoving angular
diameter distance.  
As the typical redshift of the sample is $z=0.35$, we quote our result 
for the dilation scale as $D_V(0.35)$.  
For our fiducial cosmology of $\Omega_m=0.3$, $\Omega_\Lambda=0.7$,
$h=0.7$, $D_V(0.35)=1334$ Mpc.

\begin{figure}[tb]
\plotone{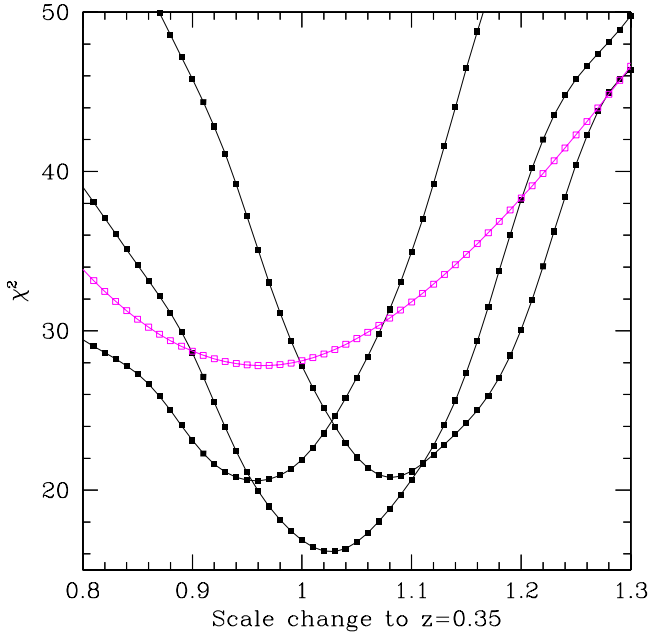}
\caption{\label{fig:test_cosm_h_all}
The $\chi^2$ values of the models as a function of the dilation
of the scale of the correlation function.  
This corresponds to altering $D_V(0.35)$ relative to
the baseline cosmology of $\Omega=0.3$, $\Omega_\Lambda=0.7$, $h=0.7$.  
Each line (save the magenta line) in the plot is a different value of 
$\Omhh$, 0.11, 0.13, and 0.15 from left to right.  
$\Obhh=0.024$ and $n=0.98$ are used in all cases. 
The amplitude of the model has been marginalized over.
The best-fit $\chi^2$ is 16.1 on 17 degrees of freedom, consistent with expectations.
The magenta line (open symbols) shows the pure CDM model with $\Omhh=0.10$; it has
a best $\chi^2$ of 27.8, which is rejected at 3.4~$\sigma$.
Note that this curve is also much broader, indicating that the lack of an
acoustic peak makes the scale less constrainable.
}
\end{figure}

We compute parameter constraints by computing $\chi^2$ (using the full covariance
matrix) for a
grid of cosmological models.  In addition to cosmological parameters of
$\Omhh$, $\Obhh$, and $n$, we include the distance scale $D_V(0.35)$
of the LRG sample and marginalize over the amplitude of the correlation
function.  Parameters such as $h$, $\Omega_m$, $\Omega_K$, and $w(z)$
are subsumed within $D_V(0.35)$. 
We assume $h=0.7$ when computing the scale at which to apply
the non-linear corrections; having set those corrections, we then dilate
the scale of the final correlation function.

The WMAP data \citep{Ben03}, as well as combinations of WMAP with large-scale structure 
\citep{Spe03,Teg03b}, 
the Lyman-alpha forest \citep{McD04,Sel04}, and big bang nucleosynthesis \citep[e.g.,][]{Bur01,Coc04},
constrain $\Obhh$ and $n$ rather well and so to begin, we hold these parameters
fixed (at 0.024 and 0.98, respectively), and consider only variations in
$\Omhh$.  In practice, the sound horizon varies only as $(\Obhh)^{-0.08}$, 
which means that the tight constraints from WMAP \citep{Spe03}
and big bang nucleosynthesis \citep{Bur01} make the 
uncertainties in the baryon density negligible.

\begin{figure}[t]
\plotone{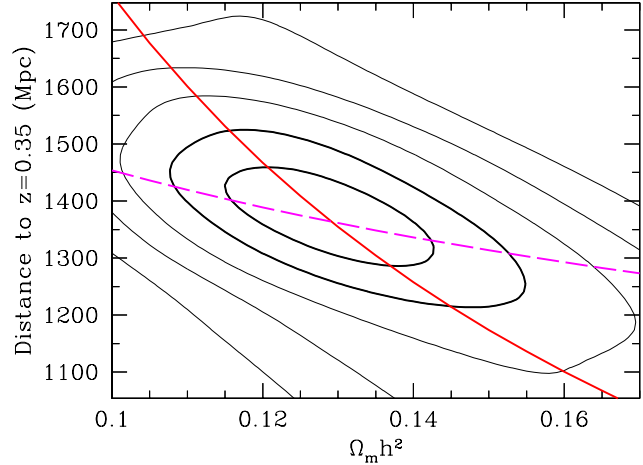}
\caption{\label{fig:lomh_gamma}
The likelihood contours of CDM models as a function of $\Omhh$
and $D_V(0.35)$.  The likelihood
has been taken to be proportional to $\exp(-\chi^2/2)$,
and contours corresponding to 1~$\sigma$ through 5~$\sigma$ for
a 2-d Gaussian have been plotted.
The one-dimensional marginalized values are $\Omhh=0.130\pm0.010$
and $D_V(0.35) = 1370\pm64$ Mpc.
We overplot lines
depicting the two major degeneracy directions.
The solid (red) line is a line of constant $\Omega_m h^2 D_V(0.35)$, 
which would be the degeneracy direction for a pure CDM model.
The dashed (magenta) line is a line of constant sound horizon, holding $\Obhh=0.024$.
The contours clearly deviate from the pure CDM degeneracy, implying 
that the peak at $100\hmpc$ is constraining the fits.
}
\end{figure}

Figure \ref{fig:test_cosm_h_all} shows $\chi^2$ as a function of the
dilation for three different values of $\Omhh$, 0.11, 0.13, and 0.15.  Scanning
across all $\Omhh$, the best-fit $\chi^2$ is 16.1 on 17 degrees of freedom
(20 data points and 3 parameters: $\Omhh$, $D_V(0.35)$, and the amplitude).
Figure \ref{fig:lomh_gamma} shows the contours of equal $\chi^2$ in $\Omhh$ 
and $D_V(0.35)$, 
corresponding to 1~$\sigma$ up to 5~$\sigma$ for a 2-dimensional Gaussian
likelihood function.  Adopting a likelihood proportional to $\exp(-\chi^2/2)$,
we project the axes to find $\Omhh=0.130\pm0.010$ and $D_V(0.35) = 1370\pm64\mpc$ (4.7\%),
where these are 1~$\sigma$ errors.

Figure \ref{fig:lomh_gamma} also contains two lines that depict the two
physical scales.  The solid line is that of constant $\Omhh D_V$, which
would place the (matter-radiation) equality scale at a constant apparent location.  This 
would be the degeneracy direction for a pure cold dark matter cosmology
and would be a line of constant $\Gamma = \Omega_m h$ were the LRG sample
at lower redshift.  The dashed line holds constant the sound horizon divided by 
the distance, which is the apparent location of the acoustic scale.
One sees that the long axis of the contours falls in between these two,
and the fact that neither direction is degenerate means that both the
equality scale and the acoustic scale have been detected.  Note that 
no information from the CMB on $\Omhh$ has been used in computing $\chi^2$,
and so our constraint on $\Omhh$ is separate from that from the CMB.

The best-fit pure CDM model has $\chi^2 = 27.8$, which means that it
is disfavored by $\Delta\chi^2=11.7$ compared to the model with $\Obhh=0.024$.
Note that we are not marginalizing over the baryon density, so these
two parameter spaces have the same number of parameters.
The baryon signature is therefore detected at 3.4~$\sigma$.
As a more stringent version of this, we find that the baryon model is
preferred by $\Delta\chi^2=8.8$ (3.0~$\sigma$) even if we only include 
data points between 60 and $180\hmpc$.
Figure \ref{fig:test_cosm_h_all} also shows $\chi^2$ for the pure-CDM
model as a function of dilation scale; one sees that the scale
constraint on such a model is a factor of two worse than the baryonic
models.  This demonstrates the importance of the acoustic scale
in our distance inferences.

\begin{table}[t]
\footnotesize
\caption{\label{tab:basic}}
\begin{center}
{\sc Summary of Parameter Constraints from LRGs\\}
\begin{tabular}{rl} 
\tableskip\tableline\tableline\tableskip
$\Omega_mh^2$ & $0.130(n/0.98)^{1.2} \pm 0.011$ \\
$D_V(0.35)$ & $1370\pm64$ Mpc (4.7\%) \\
$R_{0.35} \equiv D_V(0.35)/D_M(1089)$ & $0.0979\pm0.0036$ (3.7\%) \\
$A \equiv D_V(0.35)\sqrt{\Omega_m H_0^2}/0.35c$ & $0.469(n/0.98)^{-0.35}\pm0.017$ (3.6\%) \\
\tableskip\tableline\tableline\tableskip
\end{tabular}
\end{center}
NOTES.---%
We assume $\Omega_bh^2=0.024$ throughout, but variations permitted by WMAP
create negligible changes here.  We use $n=0.98$, but where variations by 0.1
would create 1~$\sigma$ changes, we include an approximate dependence.
The quantity $A$ is discussed in \S~\ref{sec:lowz}.
All constraints are 1~$\sigma$.  
\end{table}

As most of our distance leverage is coming from the acoustic scale, 
the most robust distance measurement we can quote is the ratio of the 
distance to $z=0.35$ to the distance to $z=1089$ (the redshift of
decoupling, Bennett et al.\ 2003).  This marginalizes
over the uncertainties in $\Omhh$ and would cancel out more exotic 
errors in the sound horizon, such as from extra relativistic species \citep{Eis04w}.  
We denote this ratio as
\beq\label{eq:R35}
R_{0.35} \equiv {D_V(0.35)\over D_M(1089)}.
\eeq
Note that the CMB measures a purely transverse distance, while the LRG
sample measures the hybrid in Equation (\ref{eq:D}).  $D_M(1089)=13700$
Mpc for the $\Om=0.3$, $\Omega_\Lambda=0.7$, $h=0.7$ cosmology \citep{Teg03b}, 
with uncertainties due to imperfect
measurement of the CMB angular acoustic scale being negligible at $<1\%$.
We find $R_{0.35} = 0.0979 \pm 0.0036$, which is a 4\% measurement of
the relative distance to $z=0.35$ and $z=1089$.
Table \ref{tab:basic} summarizes our numerical results on these basic measurements.

\begin{figure}[t]
\plotone{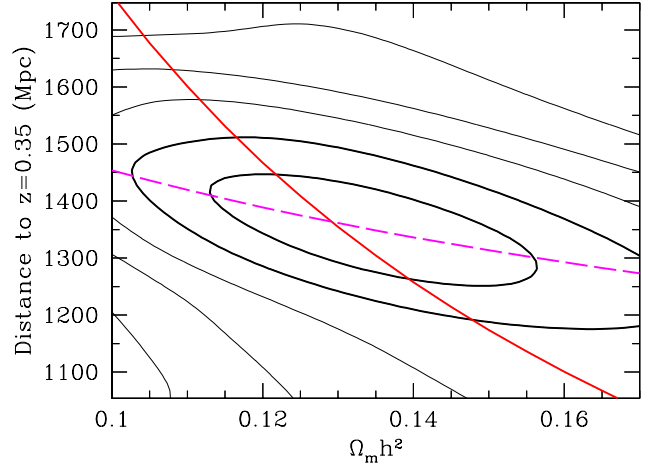}
\caption{\label{fig:lomh_gamma.r20}
As Figure \protect\ref{fig:lomh_gamma}, but now with scales below $18\hmpc$
excluded from the $\chi^2$ computation.  This leaves 18 separation bins and
15 degrees of freedom.
The contours are now obviously aligned to the line of constant sound horizon,
and the constraints in the $\Omhh$ direction are weakened by 40\%.
As Figure \protect\ref{fig:xir2_jack} would suggest,
the data at scales below $18\hmpc$ help to constrain $\Omhh$, 
twisting the contours towards the pure CDM degeneracy.
Dropping the smaller scales doesn't affect the constraint on 
$R_{0.35}$; we find $0.0973\pm0.0038$ as compared to $0.0979\pm0.0036$ before.
}
\end{figure}

\begin{figure}[t]
\plotone{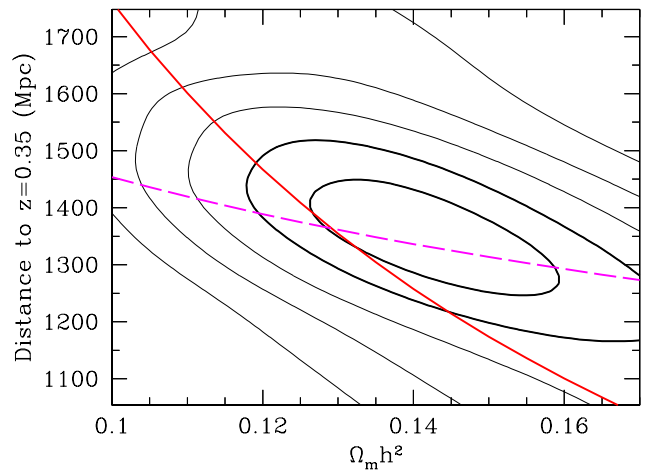}
\caption{\label{fig:lomh_gamma.n90}
As Figure \protect\ref{fig:lomh_gamma}, but now with a spectral tilt of $n=0.90$.
The best fit has $\chi^2=17.8$.
The primary effect is a shift to larger $\Omhh$,
$0.143\pm0.011$.
However, this shift occurs at essentially constant
$R_{0.35}$; we find $0.0986\pm0.0041$.  
Again, the acoustic scale robustly determines the distance, even
though the spectral tilt biases the measurement of the 
equality scale.
}
\end{figure}

To stress that the acoustic scale
is responsible for the distance constraint, we repeat our fitting having
discarded the two smallest separation bins ($10<s<18\hmpc$) from the 
correlation function.  This is shown in Figure \ref{fig:lomh_gamma.r20}.
One sees that the constraints on $\Omhh$ have degraded (to $0.136\pm0.014$),
but the contours remain well confined along the direction of constant
acoustic scale (the dashed line).  We find a distance ratio of $0.0973\pm 0.0038$,
essentially identical to what we found above, with a best $\chi^2$ of 13.7 on 15 degrees of freedom.

\begin{figure*}[tb!]
\centerline{\epsfxsize=3.5in\epsffile{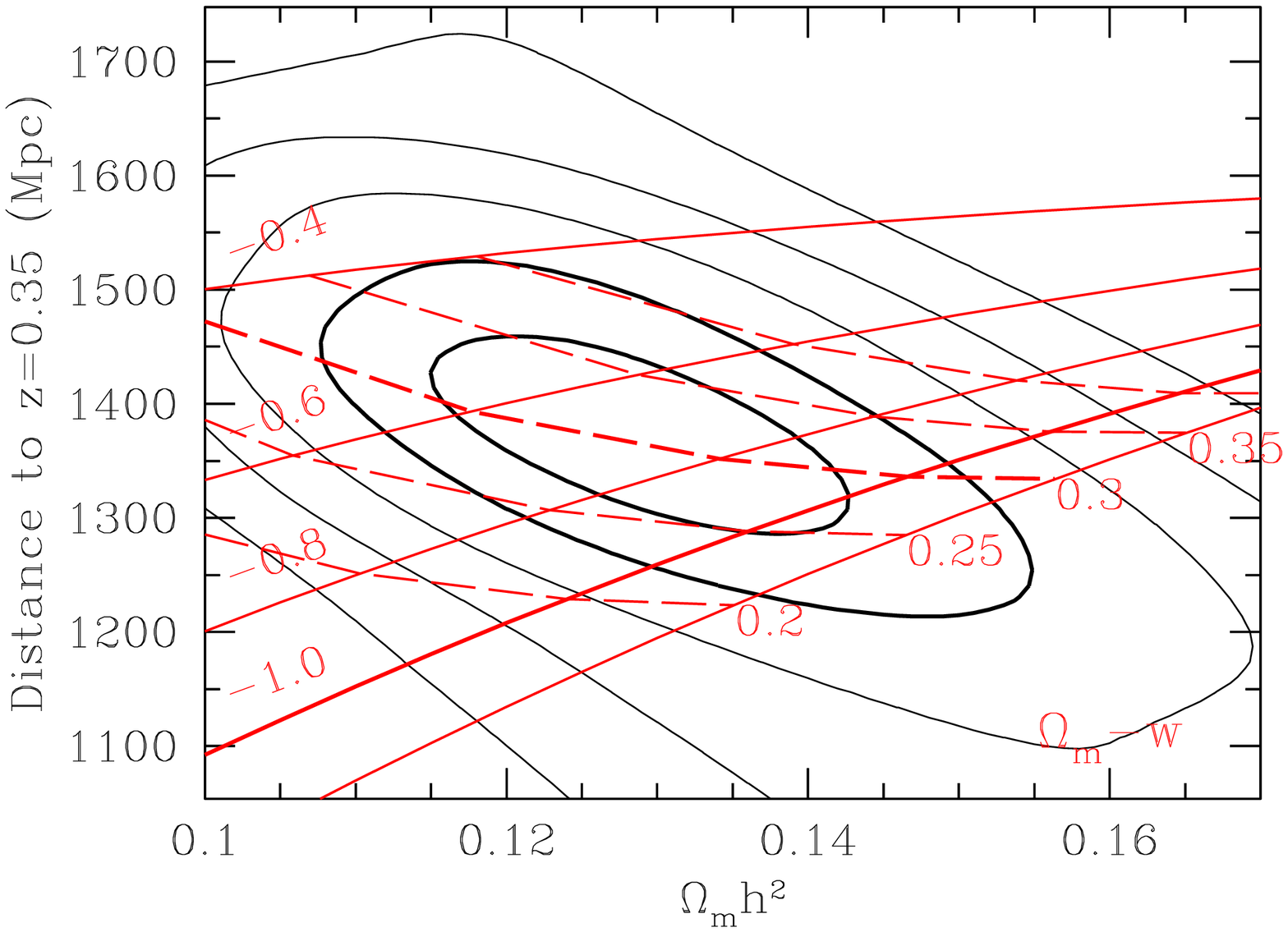}
\quad\epsfxsize=3.5in\epsffile{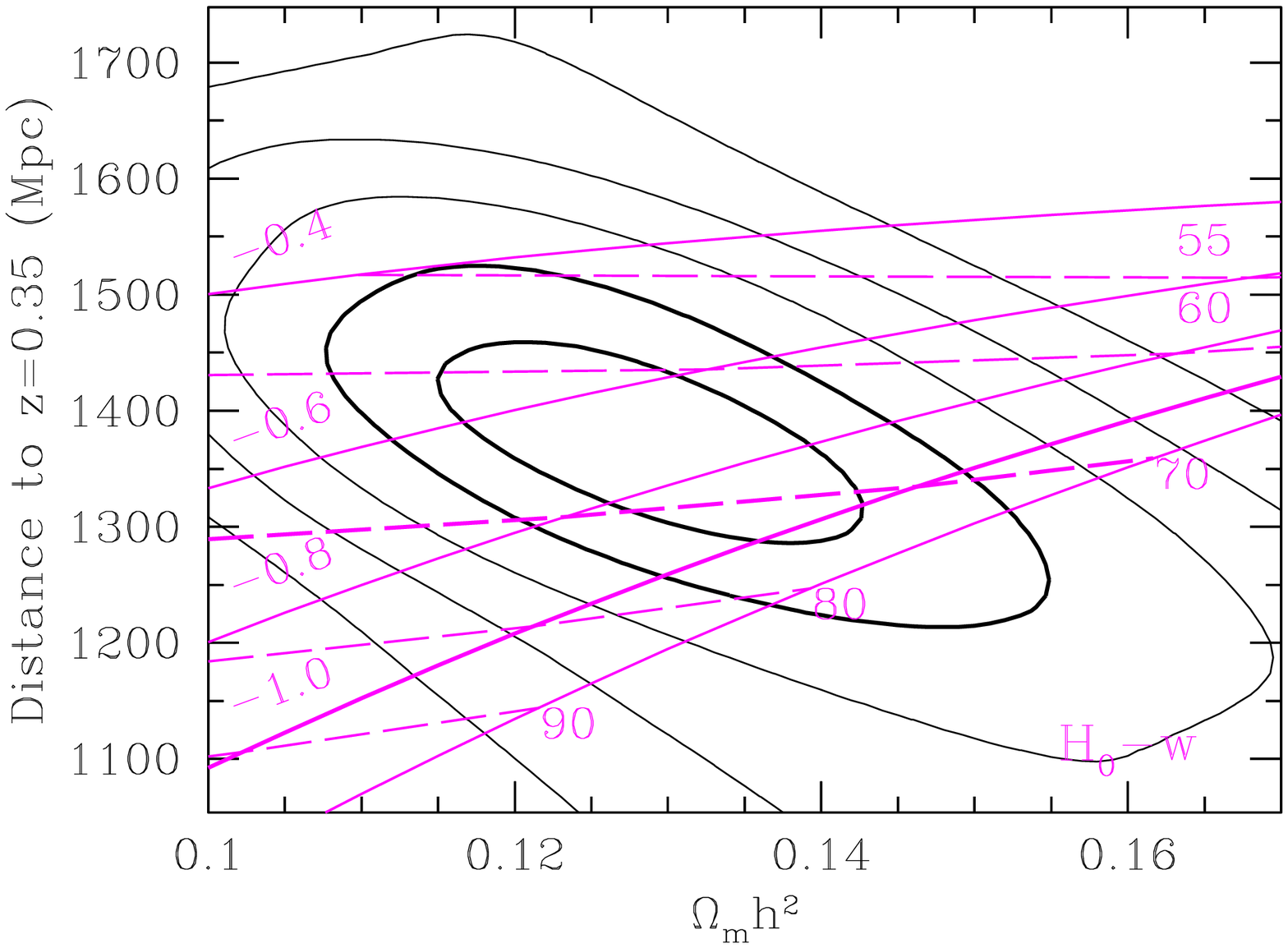}}
\caption{\label{fig:lomh_omw}
a) As Figure \protect\ref{fig:lomh_gamma}, but overplotted with
model predictions from constant $w$ flat models.  For a given
value of $\Omhh$ and $w$, the angular scale of the CMB acoustic peaks
(known to 1\%) determines $\Omega_m$ and $H_0$.  Of course, the required
$\Omega_m$ is a function of $w$ and $\Omhh$.  The solid red lines
show lines of constant $w$; the dashed lines show lines of constant
$\Omega_m$.  Our knowledge of $\Omhh$ still limits our inference of $w$.
b) As (a), but the dashed lines are now lines of constant $H_0$.
}
\end{figure*}

Varying the spectral tilt, which has a similar effect to
including massive neutrinos, is partially degenerate with $\Omhh$,
but the ratio of the distances is very stable across the plausible
range.  Repeating the fitting with $n=0.90$ changes the distance ratio
$R_{0.35}$ to $0.0986\pm0.0041$, a less than 1\% change.  
The distance itself changes by only 2\% to $D_V(0.35)=1344\pm70$.  
The change in the likelihood
contours is shown in Figure \ref{fig:lomh_gamma.n90}.  The best-fit
$\Omhh$ for $n=0.90$ is $0.143\pm0.011$, and so we approximate our $\Omhh$
constraint as $0.130(n/0.98)^{-1.2}\pm0.011$.  
While changes of order
0.08 in tilt are marginally allowed with CMB alone, they are strongly disfavored
when WMAP is combined with the Lyman-alpha forest and galaxy power spectra \citep{Sel04}.

Changing the baryon density to $\Omega_bh^2=0.030$ still yields a good fit,
$\chi^2=16.2$, but increases the inferred $\Omhh$ to $0.146\pm0.010$.
This is not surprising, because higher baryon fractions and lower
$\Omhh$ both increase the ratio of large to small-scale power.
However, $R_{0.35}$ changes only to $0.0948\pm0.0035$.  This is a 1~$\sigma$
change in $R_{0.35}$, 
whereas this baryon density change is rejected at 5~$\sigma$ by WMAP \citep{Spe03,Teg03b}.

As described in \S \ref{sec:nonlin}, our model correlation functions include two 10\%
scale-dependent corrections, the first for non-linear gravity and the
second for scale-dependent bias and redshift distortions.  Removing 
the latter changes $\Omhh$ to $0.148\pm0.011$, a 2~$\sigma$
change from the baseline.  We regard ignoring the correction as an extreme alteration.
However, even this only moves $R_{0.35}$ to $0.0985\pm0.0039$.  The best fit 
itself is worse, as $\chi^2$ increases to 19.4.

The contour plots are based on the covariance matrix derived from the
mock catalogs.  To
validate this, we consider the scatter in the best-fit model parameters
among the 10 jackknife subsamples (see discussion in \S~\ref{sec:covar}).  
The jackknifed error in $\Omhh$ is 0.011,
that in $D_V(0.35)$ is 4.6\%, and that in the distance
ratio $R_{0.35}$ is 3.2\%.  These values are 
close to those found from the $\chi^2$-based likelihood function (0.011,
4.8\%, and 3.7\%, respectively).
This justifies the likelihood contours derived from the the covariance matrix.
It also demonstrates that our results are not being driven by one unusual
region of the survey.

\subsection{Constraints on dark energy and spatial curvature}
\label{sec:wK}

For fixed values of $\Omhh$ and $\Obhh$, the angular scale of the CMB
acoustic peaks constrains the angular diameter distance to $z=1089$
to very high accuracy.  If one considers only a simple parameter space
of flat cosmologies with a cosmological constant, then this distance
depends only on one parameter, say $\Om$ or $\Omega_\Lambda$ (the two must sum to 
unity, and $H_0$ is then fixed by the value of $\Omhh$), and so the 
distance measurement constrains $\Om$, $\Omega_\Lambda$, and
$H_0$ to high precision.  
If one generalizes to larger parameter spaces,
e.g., adding an unknown dark energy equation of state $w(z)$ \citep{Tur97,Cal98}
or a non-zero
curvature, then a parameter degeneracy opens in the CMB \citep[e.g,][]{Eht98,Efs99}.  
The acoustic scale still
provides one high quality constraint in this higher dimensional space,
but the remaining directions are constrained only poorly by gravitational
lensing \citep[e.g.,][]{Sel96,Efs99}
and the Integrated Sachs-Wolfe effect on small and large angular
scales, respectively.  

With our measurement of the acoustic scale at $z=0.35$, we can add another
high quality constraint, thereby yielding good measurements on a two-dimensional
space (e.g. a constant $w\ne-1$ {\it or} a non-zero curvature).  A more
general $w(z)$ model would of course require additional input data, e.g. large-scale
structure at another redshift, supernovae distance measurements, or 
a Hubble constant measurement, etc.  

\begin{figure}[b]
\plotone{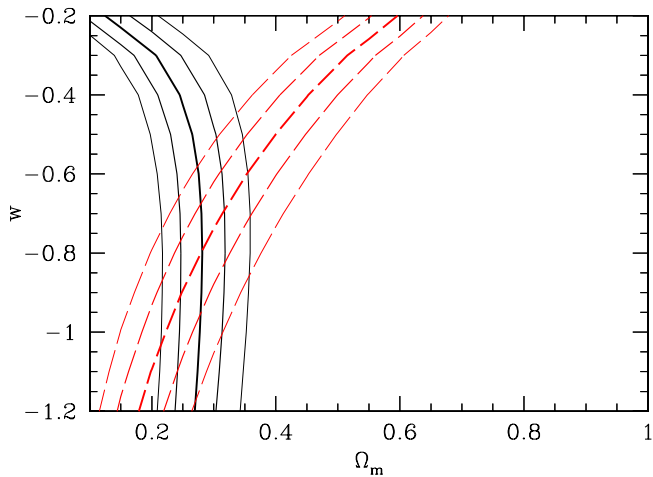}
\caption{\label{fig:DArat_cosm}
Contours in the space of $\Omega_m$ and $w$.  The solid black contours show
the lines of constant $R_{0.35}$ (from 0.090 to 0.106,
with the central value of 0.098).  The dashed red contours show the contours
of constant $\Omhh$, using the angular scale of the CMB acoustic peaks
to set $H_0$ at each ($\Omega_m$, $w$) pair.  The values of $\Omhh$
range from 0.11 to 0.15, which is the $-2$~$\sigma$
to 2~$\sigma$ range from Figure \protect\ref{fig:lomh_gamma}.
Uncertainties in the value of $\Omhh$ significantly
impact our constraints on $w$.  
}
\end{figure}

For the simple space of flat cosmologies with a constant $w\ne-1$, at
each value of $\Omhh$ and $w$, we can find the value of $\Om$ (or $H_0$)
that yields the correct angular scale of the acoustic peak and then 
use this value to predict $D_V(z=0.35)$.  In Figure \ref{fig:lomh_omw}a,
we overlay our constraints with the grid of $\Om$ and $w$ infered in 
this way.  One sees that $\Om$ is well constrained, but $w$ is not.
The reason for the latter is that $\Omhh$ is not yet known well enough.  
This is illustrated in Figure \ref{fig:DArat_cosm}.
Were $\Omhh$ known to 1\%, as is expected from the 
Planck\footnote{http://www.rssd.esa.int/index.php?project=PLANCK}
mission, then
our constraints on $w$ would actually be better than 0.1.  
Figure \ref{fig:lomh_omw}b shows the constraints with a grid of $H_0$ and $w$ overlaid.

\begin{table*}[t]
\footnotesize
\caption{\label{tab:wK}}
\begin{center}
{\sc Joint Constraints on Cosmological Parameters including CMB data \\}
\begin{tabular}{ccccccc} 
\tableskip\tableline\tableline\tableskip
& \multicolumn{2}{c}{Constant $w$ flat} 
& \multicolumn{2}{c}{$w=-1$ curved} 
& \multicolumn{2}{c}{$w=-1$ flat} \\
Parameter & WMAP+Main & +LRG & WMAP+Main & +LRG & WMAP+Main & +LRG \\
\tableskip\tableline\tableline\tableskip
$w$        & $-0.92\pm0.30$   & $-0.80\pm0.18$   & \nodata           & \nodata          & \nodata & \nodata \\
$\Omega_K$ & \nodata          & \nodata          & $-0.045\pm0.032$  & $-0.010\pm0.009$ & \nodata & \nodata \\
$\Omhh$    & $0.145\pm0.014$  & $0.135\pm0.008$  & $0.134\pm0.012$   & $0.136\pm0.008$  & $0.146\pm0.009$ & $0.142\pm0.005$ \\
$\Omega_m$ & $0.329\pm0.074$  & $0.326\pm0.037$  & $0.431\pm0.096$   & $0.306\pm0.027$  & $0.305\pm0.042$ & $0.298\pm0.025$ \\
$h$        & $0.679\pm0.100$  & $0.648\pm0.045$  & $0.569\pm0.082$   & $0.669\pm0.028$  & $0.696\pm0.033$ & $0.692\pm0.021$ \\
$n$      & $0.984\pm0.033$  & $0.983\pm0.035$  & $0.964\pm0.032$   & $0.973\pm0.030$  & $0.980\pm0.031$ & $0.963\pm0.022$ \\
\tableskip\tableline\tableline\tableskip
\end{tabular}
\end{center}
NOTES.---%
Constraints on cosmological parameters from the Markov chain analysis.
The first two data columns are for spatially flat models with constant
$w$, while the next two are for $w=-1$ models with spatial curvature.
In each case, the other parameters are $\Omhh$, $\Omega_b h^2$, $n_s$,
$h$, and the optical depth $\tau$ (which we have required to be less
than 0.3).  
A negative $\Omega_K$ means a spherical geometry.
The mean values are listed with the 1~$\sigma$ errors.
The first column in each 
set gives the constraints from \protect\citet{Teg03b} from combining 
WMAP and the SDSS Main sample.  The second column adds our LRG constraints: 
$R_{0.35} = 0.0979\pm0.036$ and and $\Omhh=0.130(n/0.98)^{-1.2}\pm0.011$.
In all cases, $\Obhh$ is constrained by the CMB to an accuracy well below 
where we would need to include variations in the LRG analysis.
\end{table*}

\citet{Teg03b} used a Markov chain analysis of the WMAP data combined with 
the SDSS Main sample galaxy power spectrum to constrain cosmological parameters.  
They found
$\Omhh=0.145\pm0.014$, $w=-0.92\pm0.30$, $\Omega_m = 0.329\pm0.074$, and $h=0.68\pm0.10$
(varying also $n$, $\Omega_b$, the optical depth $\tau$, and a linear bias).
Here we use the mean and standard deviation rather than the asymmetric
quantiles in \citet{Teg03b}, and we use a prior of $\tau<0.3$.
Adding the LRG measurement of $R_{0.35}$ and the constraint that 
$\Omhh = 0.130(n/0.98)^{-1.2}\pm 0.011$, we find
$\Omhh=0.135\pm0.008$, $w=-0.80\pm0.18$, $\Omega_m = 0.326\pm0.037$, and $h=0.648\pm0.045$.
We are ignoring the small overlap in survey region between SDSS Main and the LRG sample.
The improvements in $w$ arise primarily from the constraint on $\Omhh$, 
while the improvements in $\Omega_m$ come more from the measurement of $R_{0.35}$.
Table \ref{tab:wK} summarizes these numerical results.

\begin{figure*}[tb]
\centerline{\epsfxsize=3.5in\epsffile{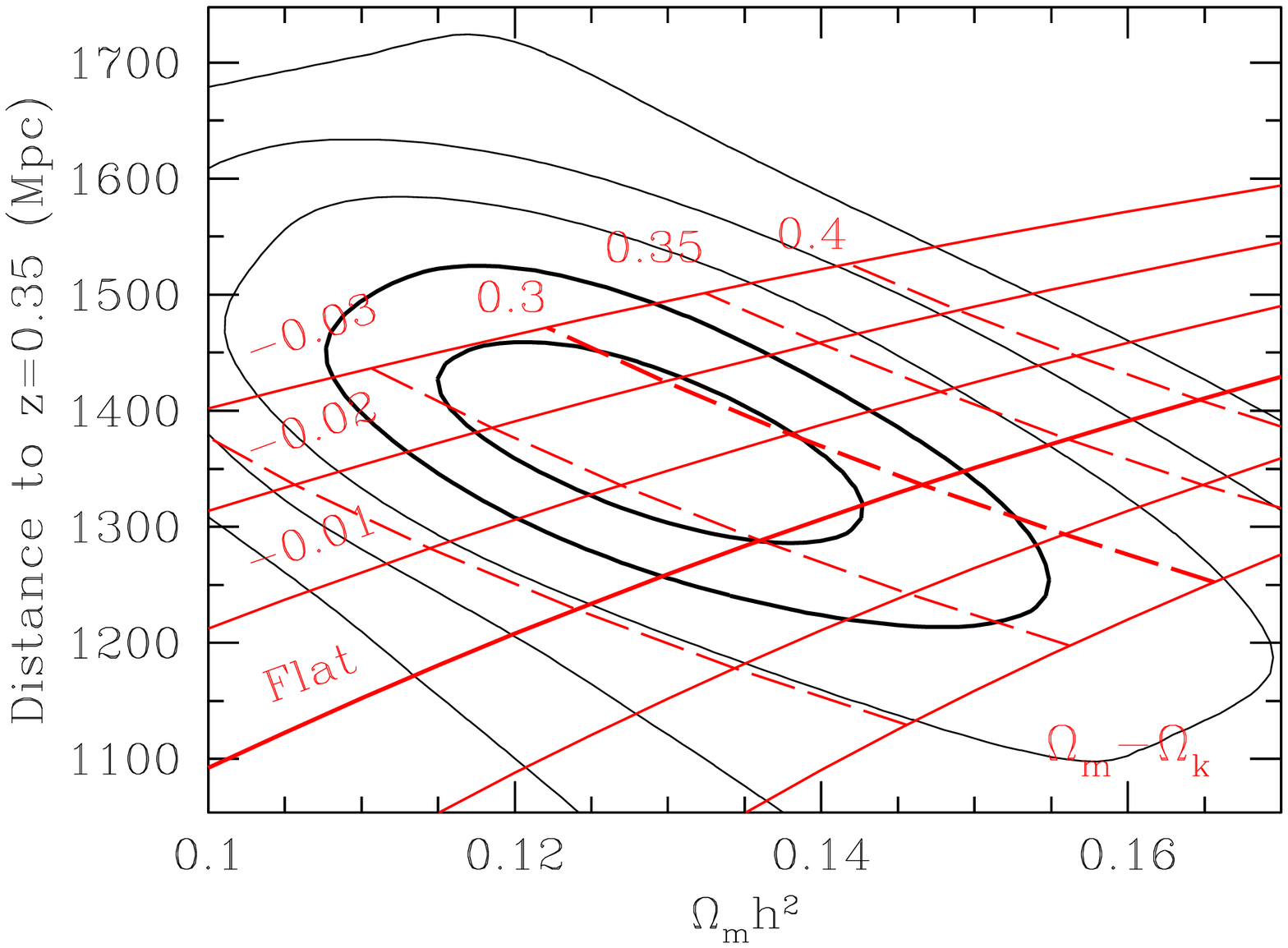}
\quad\epsfxsize=3.5in\epsffile{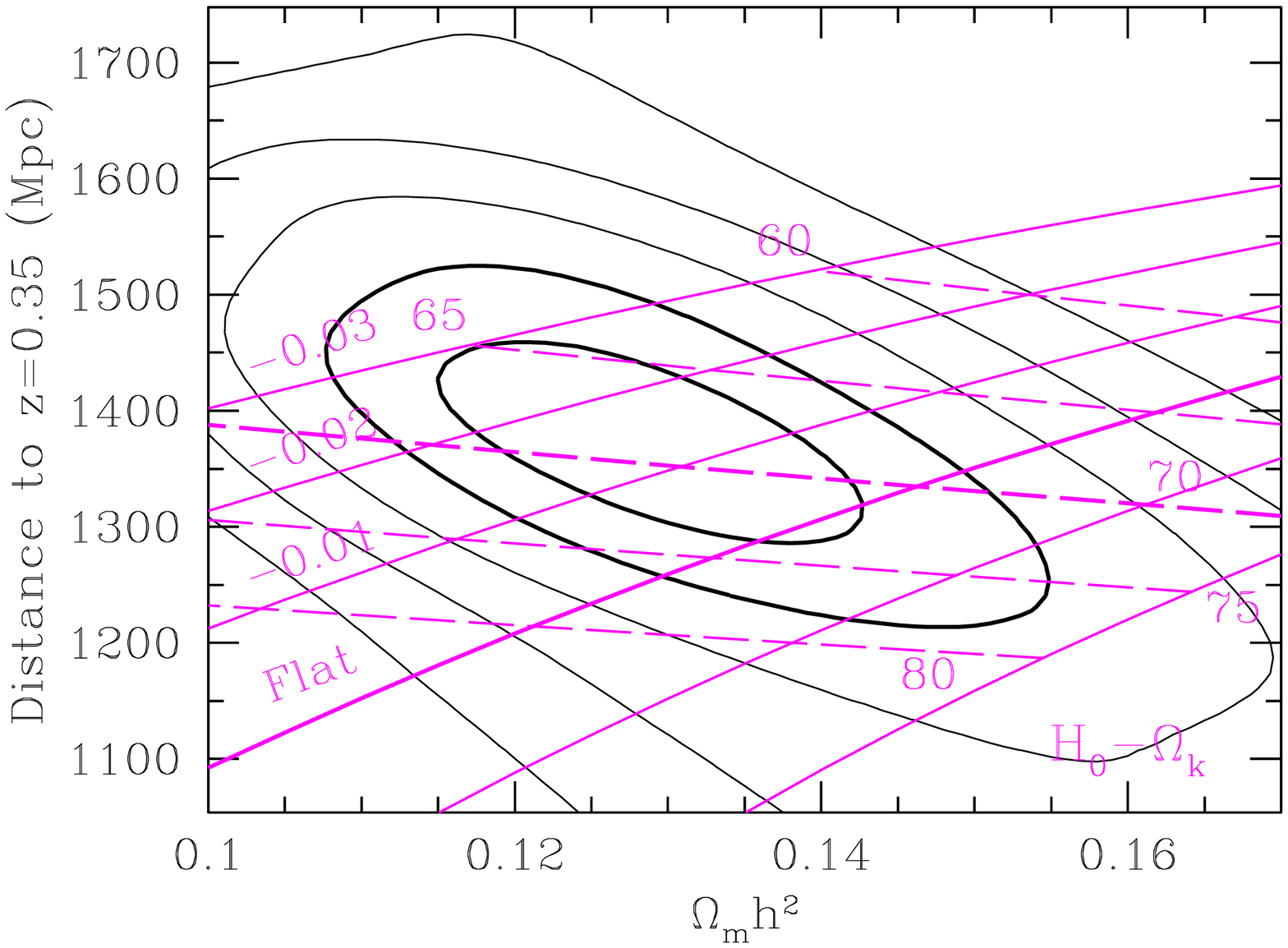}}
\caption{\label{fig:lomh_omK}
a) As Figure \protect\ref{fig:lomh_omw}, but now overplotted with
models of $w=-1$ but non-zero curvature.  The solid red lines are 
lines of constant curvature, running from $-0.03$ (closed) to $+0.02$
(open).  The dashed lines are lines of constant $\Omega_m$, from
0.20 to 0.40.
The constraints on the curvature of the universe
are superb.  This happens because we are combining
the low-redshift distance scale with the distance
to $z=1089$, by which point even tiny amounts of curvature make a 
big difference.
b) As (a), but the dashed lines are now lines of constant $H_0$.
}
\end{figure*}

It is important to remember that constant $w$ models are not necessarily
good representations of physical models of dynamical dark energy and
that forcing this parameterization can lead to bias \citep{Mao04,Bas04}.
We offer the previous analysis as a means to compare to the literature,
but we prefer our actual distance measurements in Table \ref{tab:basic}
as a model-independent set of constraints.

We next turn to the space of models with two well-specified ingredients,
namely a cosmological constant (i.e., $w=-1$) {\it and} non-zero spatial 
curvature.  
The results are in Figure \ref{fig:lomh_omK}.
Unlike the constraints on $w\ne-1$, the constraints on the spatial 
curvature are excellent, of order 1\%.  This is because the distance
to $z=1089$ is extremely sensitive to spatial curvature, such that we
get excellent performance by 
supplying a calibration of the distance scale (e.g., $H_0$ with a touch
of $\Om$) at low redshift.  Of course, this
is in accord with the conventional wisdom that the CMB constrains the universe
to be nearly flat, but our result represents a significant tightening of the angular
diameter distance degeneracy.

Using the \citet{Teg03b} Markov chain results for a $w=-1$ cosmology with 
spatial curvature, 
the SDSS Main sample $P(k)$ and WMAP produces $\Omhh = 0.134\pm 0.012$,
$\Omega_K = -0.045\pm 0.032$, $\Omega_m = 0.43\pm 0.096$, and $h=0.57\pm 0.08$.
Adding the $R_{0.35}$ constraint from the SDSS LRG results, we find
$\Omhh = 0.142\pm 0.011$,
$\Omega_K = -0.006\pm 0.011$, $\Omega_m = 0.309\pm 0.086$, and $h=0.679\pm 0.033$.
Adding the further information on $\Omhh$ drops the values to 
$\Omhh = 0.136\pm 0.008$,
$\Omega_K = -0.010\pm 0.009$, $\Omega_m = 0.306\pm 0.027$, and $h=0.669\pm 0.028$.
Hence, the essential improvement comes from the measurement of $R_{0.35}$;
with
it, we find that the universe is flat to 1\% accuracy, assuming that $w=-1$.

If we require a flat cosmology with $w=-1$, then the Markov chain analysis
from the WMAP, Main, and LRG data together
yields $\Omhh=0.142\pm0.005$, $\Om=0.296\pm0.025$, and $h=0.692\pm0.021$.
The WMAP data alone is not strongly degenerate in this parameter space \citep{Spe03},
although the galaxy data do tighten the constraints by roughly a factor of 3.
One could also read $\Om$ and $h$ directly from the ``flat'' line in 
Figure \ref{fig:lomh_omK}.  This gives $\Om=0.271\pm0.022$ and $h=0.723\pm0.017$
at a fixed spectral tilt $n=0.98$.  The difference occurs because the WMAP and SDSS Main
pull the value of $\Omhh$ higher and because the best-fit tilt in the Markov
chain is below $n=0.98$.

\subsection{Low-redshift cosmological constraints}\label{sec:lowz}

In more general dark energy models, the $R_{0.35}$ measurement will not
measure $\Omega_K$ or $w(z)$ by itself.  However, the redshift of the 
LRG sample is low enough that we can get interesting constraints focusing
on the path from $z=0$ to $z=0.35$ rather than $z=0.35$ to $z=1089$.
We note that the combination $D_V(0.35)\sqrt{\Omhh}$ has no dependence on
the Hubble constant $H_0$, since $D_V(0.35)$ is proportional to $H_0^{-1}$ (times
a function of all the $\Omega$'s and $w(z)$).  Fortuitously, this combination
is well constrained by our data, as these contours lie along the long axis 
of our constraint region.  We measure
\begin{equation}
A \equiv D_V(0.35) {\sqrt{\Omega_m H_0^2}\over 0.35 c} = 0.469 \pm 0.017 (3.6\%).
\end{equation}
This value is robust against changes in the minimum scale of data used in the fit ($0.471\pm0.021$
for $r>18\hmpc$), the spectral tilt ($0.483\pm0.018$ for $n=0.90$), and
the baryon density ($0.468\pm0.017$ for $\Omega_bh^2=0.030$).
As $A$ is independent of a dark energy model, we include its value in Table \ref{tab:basic}.

If the LRG redshift were closer to 0, then $A$ would simply be $\sqrt{\Omega_m}$.
At $z=0.35$, $A$ depends weakly on $\Omega_K$ and on $w(z)$ over
the range $0<z<0.35$.  In detail, for a flat universe and constant $w$, which we
denote as $w_0$ given the low redshift, we have
\begin{equation}
A = \sqrt{\Omega_m} E(z_1)^{-1/3} \left[ {1\over z_1} \int_0^{z_1} {dz\over E(z)}\right]^{2/3}
\end{equation}
where $E(z) = H(z)/H_0 = \left[\Omega_m(1+z)^3+\Omega_\Lambda(1+z)^{3+3w_0}\right]^{1/2}$ and $z_1=0.35$.
The generalization to curved space-times is straightforward.
While treating $w$ as a constant for all times may be
a poor model \citep{Mao04,Bas04}, 
it is a reasonable approximation for so short an interval.
In detail, $w_0$ is not the value at $z=0$ but rather some average out to $z=0.35$.

We therefore linearize the expression for $A$ in $\Omega_m$, $\Omega_K$, and $w_0$
to find
\begin{equation}\label{eq:om}
\Omega_m = 0.273 + 0.123(1+w_0)+0.137\Omega_K\pm0.025
\end{equation}
This result relies on the acoustic length scale being predicted correctly at $z\sim1000$ 
from the CMB measurement of $\Obhh$ and the LRG measurement of $\Omhh$, but it is 
independent of the angular acoustic scale in the CMB and hence makes no assumption
about $w(z)$ at $z>0.35$.  It will depend slightly
on unmarginalized parameters such as the spectral tilt, the neutrino mass, or other 
manners of altering the LRG value
of $\Omhh$.  As we demonstrated in \S \ref{sec:wK}, the CMB acoustic scale
is very sensitive to $\Omega_K$; invoking a large $|\Omega_K|$ in equation 
(\ref{eq:om}) would require large contortions in $w(z)$ to maintain the angular
location of the acoustic peaks.
The error in equation (\ref{eq:om}) is consistent with 
the error on $\Om$ in the constant $w$ Markov chain because the uncertainties
in $w_0$ increase the allowed range of $\Om$.

\section{Conclusions}

We have presented the large-scale correlation function from the SDSS
Luminous Red Galaxy sample.  This is the largest effective volume yet
surveyed by a factor of $\sim\!4$ at small wavenumber.  We find
clear evidence (3.4~$\sigma$) for the acoustic peak at $100\hmpc$ scale.
The scale and amplitude of this peak is in excellent agreement with the 
prediction from the $\Lambda$CDM interpretations of CMB data such as from WMAP.
Moreover, the broadband shape of the rest of the correlation function 
gives a measurement of the matter density $\Omhh$ that matches the CMB 
findings.

Before reviewing the quantitative conclusions, we focus on the more 
fundamental ones.  The imprint of the acoustic oscillations on the
low-redshift clustering of matter is a generic prediction of CDM cosmological
theory \citep{Pee70,Bon84,Hol89,Hu96}.  
Our detection confirms two aspects of the theory: first, that the
oscillations occur at $z\gtrsim1000$, and second that they survive the
intervening time to be detected at low redshift.  The small amplitude of the features 
requires that there exists matter at $z\sim1000$ that
does not interact with the photon-baryon fluid, i.e. dark matter.  
Fully baryonic models or those with extra interacting matter produce
much stronger acoustic signatures that would have to be erased by some
exotic later process to match low-redshift observations.  

In CDM models, large-scale fluctuations have grown since $z\sim1000$ by gravitational 
instability.  In particular, perturbation theory predicts that small perturbations
grow in a manner that leaves the Fourier modes of the density field uncoupled.
This in turn protects the narrow features such as the acoustic oscillations
as they grow.  Non-linear gravitational perturbation theories
generically predict mode coupling that would wash out the acoustic signature
\citep[e.g.][]{Fry84,Gor86,Jai94}.

Hence, the detection of low-redshift acoustic oscillations is a stark 
confirmation of the CDM theory for the growth of cosmological structure 
and the link between the CMB anisotropies and the matter perturbations.
While the agreement between recent results on the broadband shape of the 
matter power spectrum \citep{Efs02,Teg03a,McD04}, the galaxy three-point
correlation function \citep[e.g.,][]{Fel01}, and the inferences from
the CMB \citep[e.g.,][]{Spe03} 
were certainly compelling on this point, we regard the
acoustic signature as a smoking gun, as its narrowness in real space
would be difficult to mimic in alternative models of structure formation.
This detection confirms the applicability of linear cosmological perturbation 
theory on large scales and across a factor of 800 in cosmic expansion.

The narrowness of the acoustic peak in real space offers an opportunity 
to measure distances to higher redshifts \citep{Eht98,Eis03,Bla03}.  
It is worth noting that
this is a circumstance where a given improvement in signal-to-noise ratio in the
clustering statistic makes a super-linear improvement in the 
distance constraint.  One can draw an analogy to the determination of the
redshift of a galaxy with an emission line.  A factor of two in signal-to-noise
ratio can make the difference between detecting the line, and hence constraining
the redshift to very high precision, and not detecting it and having to rely 
on the spectral shape for a low-precision photometric redshift.  In the
case of large-scale clustering, the acoustic scale is not as narrow and so
the improvement is less dramatic, but we clearly benefit at the factor
of two level from using the acoustic scale rather than the broadband
shape of the correlation function (i.e. the equality scale).

In the LRG sample, we measure the acoustic scale to just better than 4\%
precision (1~$\sigma$).  Comparing this scale to the angular scale of the
CMB anisotropies gives the distance ratio $R_{0.35}=D_V(0.35)/D_M(1089)=
0.0979\pm0.0036$, where $D_V(0.35)$ is defined in equation (\ref{eq:D}).
This distance ratio is robust against changes in the broadband clustering 
signal such as via the spectral tilt and against variations in our analysis.
It is also robust against certain kinds of exotica, such as adding additional
relativistic energy to the universe \citep{Eis04w}.
It does rely on the well-understood linear perturbation theory of the 
recombination epoch to relate the perturbations in the photons to those
in the matter.  Given this theory, we have measured
the relative distance between two radically different redshifts using
a purely geometric method and the same physical mechanism.

This distance ratio is consistent with the familiar cosmological constant
cosmology.  It is grossly inconsistent with the Einstein-de Sitter ($\Omega_m=1$)
model, which predicts $R_{0.35}=0.133$ (nominally 10~$\sigma$).
A model lacking dark energy would require $\Om=0.70$ with $\Omega_K=0.30$
to match the distance ratio.  This would require $h=0.90$ and $\Omhh=0.57$
to match the CMB peak location, implying an age of 8 Gyr.  
This is in complete disagreement with
the observed shape of the CMB anisotropy spectrum, the galaxy correlation 
function (including these LRG data), the cluster baryon fraction \citep{Whi93}, 
the observed value of $H_0$ \citep{Fre01}, and the age of old stars 
\citep[][and references therein]{Kra03}, 
as well as other cosmological measurements.
Hence, our measurement provides geometric evidence for dark energy.

The size of the acoustic scale is predicted by very simple physics,
namely the comoving distance that a sound wave can travel between
the generation of the perturbations and the epoch of recombination.
In the standard cosmological model, this depends only on the matter
density $\Omhh$ and baryon density $\Obhh$.  The uncertainties on $\Obhh$
from CMB and Big Bang nucleosynthesis are small, contributing $<2\%$ 
to the error on the acoustic scale.
However, current uncertainties in $\Omhh$ are large enough that we need
to track the covariance of $\Omhh$ with our distance inferences.

Because the acoustic scale is detected, the LRG data alone constrain the
equality scale and matter density, i.e. we measure $\Omhh$ rather than the
more familiar $\Gamma = \Omega_m h$.  With our baseline method and $\Obhh=0.024$,
we find $\Omhh = 0.130(n/0.98)^{1.2}\pm 0.010$.  This precision is similar to that from 
current CMB measurements \citep{Spe03}.  Importantly, the LRG value agrees 
with the CMB value and with the inference from the clustering of the 
lower redshift SDSS Main galaxy sample \citep{Teg03a}, a remarkable 
cosmological consistency test.  Because the formal precision of the LRGs is as good
as the other measurements, we choose to use only the LRGs in our fitting.
Adding the WMAP information on $\Omhh$ as an external prior would improve 
the quantitative constraints only slightly, and so we leave it as a cross-check.  
We expect our knowledge of 
$\Omhh$ to improve rapidly in the coming years both from the CMB, with
additional WMAP data and smaller angle ground-based observations, and from 
large-scale structure, e.g. with improved modeling of scales below $10\hmpc$
and the continued data collection for the SDSS LRG sample.

Using the LRG value for $\Omhh$, we find the distance to $z=0.35$ to be
$D_V(0.35) = 1370\pm 64$ Mpc, a 5\% measurement.  Were this at $z\sim0$,
we would have a measurement of $H_0$ and $\Omega_m$, but at $z=0.35$ dark
energy and curvature do matter.  The combination $D_V(0.35)\sqrt{\Omega_m
H_0^2}/0.35c$ is measured to 4\% precision and is independent of $H_0$.
From this, we infer $\Omega_m=0.273+0.123(1+w_0)+0.137\Omega_K\pm0.025$, 
where this $w_0$ is the effective value in the range $0<z<0.35$.

Combining with the CMB acoustic scale, we put constraints on more restricted
models, either constant $w$ or $w=-1$ plus curvature.  We find that our
$w$ leverage is roughly $0.2$.  Improvements in knowledge of $\Omhh$
will help significantly.  Our leverage on spatial curvature is exquisite:
we measure $\Omega_K = -0.010\pm0.009$.  Of course, this is a manifestation
of the well-known sensitivity of the CMB to spatial curvature, but we are
breaking the angular diameter distance degeneracy with the best precision
to date and with a distance ratio that relies on the same physics as the CMB.

It is important to note that because the low-redshift acoustic oscillation
method measures distances that can be quantitatively compared to those
from the CMB, the method retains sensitivity to phenomena that have more
effect at higher redshift, such as curvature.  Relative distance methods,
such as supernovae, can only constrain the Hubble relation out to the
maximum redshift of the sample, but absolute methods probe both above 
and below that redshift by using both $z=0$ and the CMB as comparison points.

It is interesting to compare the provenance of the absolute distance measurements 
offered by the acoustic scale to those of the classical measurements of $H_0$
\citep[e.g.][]{Fre01}.
Once established at $z>1000$, the acoustic scale can be used at low redshift
as a standard ruler on equal footing to any other.  The issue of course is 
that the early Universe ($10^3<z<10^5$) is a remote place to calibrate one's ruler.
There are assumptions about the relativistic energy density, the adiabatic
nature of the perturbations, the early generation ($z\gtrsim10^5$) of the 
perturbations, and the absence of particle decays at $z\lesssim10^5$ \citep{Eis04w}.
Many possible alterations create glaring deviations in the CMB anisotropies 
\citep[e.g.][]{Moo04}.
Others, such as small alterations to the relativistic density, are more subtle, 
at least with present data.  Our sense is that altering the acoustic scale
so as to misestimate $H_0$ will require some interesting piece of new fundamental
physics.  The future of direct $H_0$ studies may be as a probe of high-redshift
particle physics!

This detection of the acoustic peak at low redshift dramatically
confirms several basic assumptions of cosmological structure formation theory,
but it also points the way to a new application of large-scale structure
surveys for the study of dark energy.  Survey volumes of order $1\hgpcC$
offer a reliable standard ruler, whose measurement across a range of
redshifts can determine $H(z)$ and $D_A(z)$ robustly to percent-level accuracy
\citep{Bla03,Hu03,Seo03}.  Indeed, the available precision improves
at higher redshift because the acoustic peak is less broadened by
non-linear structure formation.
The observational challenge is to execute these large, wide-field surveys at $z>0.5$.

\bigskip

We thank Scott Dodelson, Sebastian Jester, and Ravi Sheth
for useful comments.  DJE and DWH thank the Laboratoire de Physique Th\'eorique, 
Universit\'e de Paris XI for hospitality.
DJE, IZ, and HS are supported by grant AST-0098577 and AST-0407200 from the 
National Science Foundation.  
DJE is further supported by an Alfred P. Sloan Research Fellowship.
DWH and MRB were supported by NSF PHY-0101738 and NASA NAG5-11669.
Our PTHalos mock catalogs were created on the NYU Beowulf cluster supported by
NSF grant PHY-0116590.

Funding for the creation and distribution of the SDSS Archive has
been provided by the Alfred P. Sloan Foundation, the Participating
Institutions, the National Aeronautics and Space Administration,
the National Science Foundation, the U.S. Department of Energy, the
Japanese Monbukagakusho, and the Max Planck Society. The SDSS Web site
is http://www.sdss.org/.

The SDSS is managed by the Astrophysical Research Consortium (ARC)
for the Participating Institutions. The Participating Institutions are
The University of Chicago, Fermilab, the Institute for Advanced Study,
the Japan Participation Group, The Johns Hopkins University, the Korean
Scientist Group, Los Alamos
National Laboratory, the Max-Planck-Institute for Astronomy (MPIA), the
Max-Planck-Institute for Astrophysics (MPA), New Mexico State University,
University of Pittsburgh, University of Portsmouth, Princeton University, 
the United States Naval Observatory, and the University of Washington.

{}

\end{document}